\documentclass[%
 reprint,
nofootinbib,
 amsmath,amssymb,
 aps,
]{revtex4-1}

\usepackage{color}
\usepackage{graphicx}
\usepackage{dcolumn}
\usepackage{bm}
\usepackage{hyperref}
\usepackage{tcolorbox}

\def\bea{\begin{eqnarray}}
\def\eea{\end{eqnarray}}
\def\be{\begin{equation}}
\def\ee{\end{equation}}
\def\ba{\begin{array}}
\def\ea{\end{array}}

\def\nn{\nonumber}

\def\k{{\bf k}}

\def\R{{\mathcal R}}
\def\F{{\mathcal F}}

\def\P{{\mathcal P}}

\def\Y{Y}
\def\X{X}

\def\equationautorefname~#1\null{Eq.~#1\null}

\begin{document}

\centerline{DESY-19-121}

\title{Origin of ultra-light fields during inflation and their suppressed non-Gaussianity}

\author{Ana Ach\'ucarro$^{a,b}$, Gonzalo A. Palma$^{c}$, Dong-Gang Wang$^{a,d}$, and Yvette Welling$^{a,d,e}$}

\affiliation{%
\it $^{a}$Lorentz Institute for Theoretical Physics, Leiden University, 2333CA Leiden, The Netherlands \\
$^{b}$Department of Theoretical Physics, University of the Basque Country, 48080 Bilbao, Spain\\
$^{c}$Grupo de Cosmolog\'ia y Astrof\'isica Te\'orica, Departamento de F\'{i}sica, FCFM,\\ \mbox{Universidad de Chile}, Blanco Encalada 2008, Santiago, Chile. \\
$^{d}$Leiden Observatory, Leiden University, 2300 RA Leiden, The Netherlands\\
$^{e}$Deutsches Elektronen-Synchrotron DESY, Notkestra{\ss}e 85, 22607 Hamburg, Germany
}%

\date{\today}

\begin{abstract}

We study the structure of multi-field inflation models where the primordial curvature perturbation is able to vigorously interact with an ultra-light isocurvature field -- a massless fluctuation orthogonal to the background inflationary trajectory in field space. We identify a class of inflationary models where ultra-light fields can emerge as a consequence of an underlying ``scaling transformation" that rescales the entire system's action and keeps the classical equations of motion invariant. This
scaling invariance
ensures the existence of an ultra-light fluctuation that freezes after horizon crossing. If the inflationary trajectory is misaligned with respect to the scaling symmetry direction, then the isocurvature field is proportional to this ultra-light field, and becomes massless. In addition, we find that even if the isocurvature field interacts strongly with the curvature perturbation --transferring its own statistics to the curvature perturbation-- it is unable to induce large non-Gaussianity. The reason is simply that the same mechanism ensuring a suppressed mass for the isocurvature field is also responsible for suppressing its self-interactions. As a result, in models with light isocurvature fields the bispectrum is generally expected to be slow-roll suppressed, but with a squeezed limit that differs from Maldacena's consistency relation.

\end{abstract}


\maketitle


\section{\label{sec:intro}Introduction}

What are the general conditions leading to primordial non-Gaussianity in multi-field models of inflation? A distinctive feature of multi-field inflation is that isocurvature fields (fluctuations orthogonal to the background inflationary trajectory in field space) can transfer their statistics to the primordial curvature perturbation~\cite{Achucarro:2016fby, Achucarro:2019pux}. This transfer can in principle enhance the generation of primordial non-Gaussianities as long as the isocurvature field experiences sizable self-interactions~\cite{Enqvist:2004bk, Lyth:2005fi, Seery:2005gb, Rigopoulos:2005ae, Alabidi:2005qi, Battefeld:2007en, Choi:2007su, Byrnes:2008wi, Byrnes:2009qy, Battefeld:2009ym, Chen:2009we, Chen:2009zp, Elliston:2011et, Mulryne:2011ni, McAllister:2012am, Byrnes:2012sc, Baumann:2011nk, Arkani-Hamed:2015bza, Lee:2016vti, Chen:2018uul, Chen:2018brw} (see also some recent works on non-Gaussianities in multi-field inflation with a curved field space~\cite{Garcia-Saenz:2018ifx, Fumagalli:2019noh, Achucarro:2019mea, Welling:2019bib, Garcia-Saenz:2019njm}). Understanding in detail this process would allow us to distinguish multi-field models from single field models~\cite{Guth:1980zm, Starobinsky:1980te, Linde:1981mu, Albrecht:1982wi, Mukhanov:1981xt} in future surveys aimed at characterizing non-Gaussian patterns in the primordial distribution of curvature perturbations.\\

At linear order, the interaction between the curvature field $\R$ and other scalar degrees of freedom depend on the interplay of two key parameters: The entropy mass $\mu$ of the isocurvature field $\sigma$ and the turning rate $\Omega$ of the trajectory~\cite{Gordon:2000hv, GrootNibbelink:2000vx, GrootNibbelink:2001qt, Achucarro:2010jv, Achucarro:2010da}. This can be seen directly in the quadratic action of a general two-field inflationary model:
\bea
S &=& \int d^4 x a^3  \bigg[  \epsilon \left(\dot \R  - \frac{2 \Omega}{\sqrt{2 \epsilon}}   \sigma\right)^2 -\frac{\epsilon}{a^2} (\nabla \R)^2 \nn \\
 &&  + \frac{1}{2}  \dot \sigma^2 - \frac{1}{2 a^2} (\nabla \sigma)^2  - \frac{1}{2}  \mu^2 \sigma^2 \bigg] , \label{action-quadratic-intro}
\eea
where $a$ is the scale factor and $\epsilon$ is the usual first slow-roll parameter.  As emphasized in~\cite{Achucarro:2010da}, a non-vanishing turning rate can be interpreted geometrically as sizing how curved (non-geodesic) the inflationary trajectory is in field space, and so one would expect $\Omega \neq 0$ to be a generic characteristic of multi-field dynamics. The interaction strength coupling together curvature and isocurvature fields is proportional to $\Omega / H$, but the effect of this interaction is limited to a period of time determined by $\mu$. This is because, for a given wavelength, the amplitude of the isocurvature perturbation decays after horizon crossing. This decay is proportional to $\exp (- \frac{\mu^2}{3H^2}  N )$ if $\mu \lesssim 3 H /2$, or proportional to $\exp (-  N / 3 )$, if $\mu \gtrsim 3 H /2$, where $N$ is the number of $e$folds after horizon crossing. Thus, if the isocurvature field has a mass of order $H$ or larger, the amplitude of the isocurvature mode vanishes after a few $e$folds, and the interaction between the two modes becomes negligible.\\

If the entropy mass $\mu$ is much smaller than $H$ (the ultra-light limit), the amplitude of the isocurvature modes freezes, and the interaction between the curvature and isocurvature modes persists after horizon crossing. At linear order, this implies that the amplitude of the curvature perturbation grows after horizon crossing, sourced by the frozen isocurvature fluctuation~\cite{Achucarro:2016fby}. As a result, the curvature perturbation inherits more efficiently the statistics of the isocurvature perturbation, as determined by its self-interactions. In principle one may expect large levels of local non-Gaussianity.\\

However, currently known scenarios where ultra-light isocurvature fields emerge show negligible levels of non-Gaussianity. A recent example is offered by a class of models called {\it shift-symmetric orbital inflation}~\cite{Achucarro:2019pux}. In these models, the isocurvature field remains light, and it can interact strongly with the curvature perturbation. Nevertheless, one finds that the amount of non-Gaussianity parametrized by the $f_{\rm NL}$ parameter is suppressed by slow roll parameters, in a way similar to single field models of inflation. \\

The purpose of this article is to examine, more closely, the ultra-light regime of multi-field inflation, paying special attention to the generation of local non-Gaussianity and the possible mechanisms ensuring that the isocurvature perturbation remains light ($\mu^2 \ll H^2$). General field theory arguments suggest that $\mu$ must be of order $H$ or much larger, unless a symmetry ensures its smallness. In Ref.~\cite{Achucarro:2016fby} it was pointed out that, at linear order, a symmetry enforcing
a vanishing mass, but allowing a non-vanishing $\Omega$ is given by
\bea
\sigma \to \sigma + c , \qquad \dot \R \to \dot \R + \frac{2 \Omega}{\sqrt{2 \epsilon}} c ,  \label{sym_intro}
\eea
for a constant parameter $c$. The question we wish to answer then is:  Is there a  symmetry at a more fundamental level~\footnote{At the level of the UV theory from where the action~(\ref{action-quadratic-intro}) for the fluctuations is derived.} ensuring that the action for the fluctuations $\R$ and $\sigma$ remains invariant under the transformation (\ref{sym_intro})? As we shall see, a partial answer to this question is that there is at least a class of models where the emergence of this symmetry can be attributed to a rescaling of the full action (scalars plus gravity) under a non-trivial simultaneous transformation of fields and coordinates. \\

\subsection{Summary of results}
For convenience of the reader we now summarize our findings and refer to the main formulas in our paper.\\

In Section~\ref{sec:dualitytransformation} we introduce a class of two-field models characterized by admitting a similarity transformation whereby the full action of the theory (including gravity) rescales by a single factor. This transformation arises from an extension of an isometry of the hyperbolic kinetic term
\begin{equation}
 -2K = 4R_0^2 \frac{\partial T \partial \bar{T}}{(T+\bar{T})^2} = (\partial Y)^2 + e^{2Y/R_0} (\partial X)^2\ ,
\end{equation}
with $T = e^{-\frac{Y}{R_0}}+ i \frac{X}{R_0}$, under which the potential rescales. The resulting similarity transformation leaves invariant
the equations of motion determining the background inflationary trajectory. \\

Crucially, the transformation allows one to identify the existence of an ultra-light field $\F$, which is defined as the fluctuation along the scaling direction. It turns out that the field $\F$ can be identified with $\R$ or $\sigma$ depending on which inflationary attractor the system chooses. More precisely, in Section~\ref{sec:attractors} we identify two classes of inflationary attractors in two-field systems with this similarity:
\begin{itemize}
\item Trajectories aligned along the scaling transformation direction (symmetry probing solutions \cite{Nicolis:2011pv}): These are trajectories whose evolution probes the scaling transformation. In this case, the $\F$-field coincides with the curvature perturbation $\R$. The isocurvature field $\sigma$ represents a fluctuation orthogonal to $\F$, and therefore its mass is not protected to remain small. An example model is given by the potential
\begin{equation}
 V \propto X^2 + X^{2.5} e^{-Y} \ ,
\end{equation}
with $R_0 = 0.5$.
\item Trajectories misaligned with respect to the direction of the scaling transformation:
In this case, the fields $\sigma$ and $\F$ become proportional to each other, and the isocurvature perturbation $\sigma$ inherits the shift symmetry respected by $\F$, which leads to a realization of (\ref{sym_intro}). An example model is given by the potential
\begin{equation}
 V \propto X^{2\beta}\ .
\end{equation}
\end{itemize}
In other words, we find that the second class of attractors allow configurations where the isocurvature fields can remain light and still interact vigorously with the curvature perturbation for most of the inflationary history. This we confirm by explicit computation in Section~\ref{sec:physicalperturbations} and in the remaining part of the paper we therefore focus on the second class of attractors. \\

On the other hand, (\ref{sym_intro}) will inevitably have consequences for non-Gaussianity. Indeed, as we will see in Section~\ref{sec:scalingtransformation} and \ref{sec:squeezedlimitbispectrum}, invariance of the action for the fluctuations
 under the {non-linear version of the} transformation (\ref{sym_intro}) will keep self interactions of the isocurvature fields suppressed, forbidding the appearance of large non-Gaussianity. In Section~\ref{sec:scalingtransformation} we analyse the consequences of the similarity on the ultra-light perturbations. First, the correlation functions of $\F$ are found to be invariant under this transformation. Furthermore, we apply the scaling property of the system to compute the squeezed limit of the bispectrum of the ultra-light field $\F$. The argument used to derive this limit (based on the background-wave method) is similar to that encountered in the derivation of Maldacena's consistency relation~\cite{Maldacena:2002vr, Creminelli:2004yq, Cheung:2007sv, Bravo:2017wyw} in single field models of inflation. It is based on the observation that the long modes of massless fields cannot be distinguished from the background, hence leading to a modulation of the power spectrum of short-wavelength modes that can be used to infer the amplitude of the squeezed limit of the bispectrum, which is found to be given as
 \bea
&&\langle \mathcal F_S (\k_1) \mathcal F_S (\k_2) \mathcal F_L (\k_3)  \rangle_{\rm sq} = \nn \\
&& (2 \pi)^3  \delta^{(3)}(\sum_i \k_i) (1-n_{\F}) P_{\F}  (k_L) P_{\F}  (k_s) ,
\label{eqn:wardidentity-intro}
\eea
which corresponds to equation \eqref{eqn:wardidentity}, the main result of Section~\ref{sec:scalingtransformation}. Here $n_{\F}$ is the spectral index of  $P_{\F}$ which is found to be slow roll suppressed (for reasons analogous to those encountered in the suppression of the spectral index in single-field slow-roll inflation). \\

In Section \ref{sec:squeezedlimitbispectrum} we perform a gauge transformation to go to the comoving gauge, and deduce the squeezed bispectrum of curvature perturbations. Our main results are given in \eqref{eqn:squeezedlimitbispectrum1} and \eqref{eqn:squeezedlimitbispectrum2}. We find that, even though the ultra-light limit corresponds to a regime where the multi-field nature of inflation is accentuated (because $\Omega \neq 0$), the final predictions are similar to those of single-field inflation, although they differ in the details. In particular, because of (\ref{eqn:wardidentity-intro}) the squeezed limit of the bispectrum is found to be suppressed, but is still given by a distinctive combination of slow-roll parameters --different from the single-field case-- that future surveys could constrain~\cite{Cabass:2018jgj}. \\

Understanding the point that a small entropy mass along the entire inflationary history {can be related to} small self-interactions, one can have a more general understanding on how primordial non-Gaussianity can be enhanced in general. The existence of self-interactions necessarily will come together with a non-vanishing mass, and so, after horizon crossing, there will be a limited amount of time in which the non-Gaussianity produced by this self-interaction can be transferred to the curvature perturbation. \\

\section{Scaling transformation}\label{sec:dualitytransformation}

A period of inflation in which the isocurvature field remains light can be sustained if there is a mechanism that protects the mass from acquiring large values. We will show that one way to realize a (nearly) vanishing entropy mass is by assuming a symmetry that relates different background solutions to each other, where one of the fields shifts by a constant. For the purpose of the present discussion, it will be sufficient for us to consider a two-field model of inflation with a hyperbolic field space. The kinetic term can be written in the following form
\begin{equation}
K = -\frac{1}{2}(\partial \Y)^2 - \frac{1}{2} e^{2 \Y/R_0}(\partial\X)^2.
\label{eqn:K}
\end{equation}
Here $R_0$ sets the negative curvature of the field space as $\mathbb{R} = - 2 / R_0^2$. Recall that hyperbolic spaces are maximally symmetric. A particular consequence of this is that the kinetic term is invariant under the following reparametrization of the fields
\begin{align}
\Y(x)&\to \Y'(x) = \Y(x) + \Lambda c,   \label{eqn:init-transformation-1} \\
\X(x)&\to \X'(x) = e^{-c \Lambda/R_0}\X(x),    \label{eqn:init-transformation-2}
\end{align}
where $\Lambda$ is a given mass scale, and $c$ is an arbitrary dimensionless constant parametrizing the redefinition of fields.
Inflation requires  a non-trivial potential, which necessarily breaks the previous symmetry on the level of the action. However, we can preserve it \textit{as a symmetry of the equations of motion}, if the potential satisfies the condition
\begin{equation}
 \X V_\X -R_0 V_\Y = 2 \beta V, \qquad  \beta \equiv\frac{R_0}{\Lambda} ,
\label{eqn:potentialidentity}
\end{equation}
where $V_X = \partial_X V$ and $V_Y = \partial_Y V$. An example of a multi-field potential with this property is given by
\begin{equation} \label{potential}
V (\X , \Y) =  X^{2 \beta} G (X e^{Y / R_0})  .
\end{equation}
where $G$ is an arbitrary function of the particular combination $X e^{Y / R_0}$. The function $G$ and the coefficient $\beta$ may be tuned appropriately in order to achieve inflation with the right characteristics (\emph{i.e.} spectral index and tensor to scalar ratio compatible with observations). Notice that a simple monomial potential of the form $V(X) \propto X^{2 \beta}$ (where $\beta$ is an arbitrary real number) already satisfies the constraint ~(\ref{eqn:potentialidentity}). The potential in (\ref{potential}) should not be regarded to be valid everywhere in field space. Towards the end of inflation, where the system has to enter a reheating phase, new operators in $V$ must appear in order to break the relation in Eq.~(\ref{eqn:potentialidentity}), allowing the system to end inflation. \\

The action obtained by putting together $K$ and $V$ is
\be
S = \int d^4 x \sqrt{- g} \left( \frac{M_{\rm Pl}^2}{2} R + K - V \right) , \label{basic_action}
\ee
where $R$ is the Ricci scalar and $g$ represents the determinant of the spacetime metric $g_{\mu \nu}(x)$. Because of the presence of $V$, it should be clear that this action is not invariant under the set of transformations (\ref{eqn:init-transformation-1}) and (\ref{eqn:init-transformation-2}) under which $K$ is invariant. However, we may extend these transformations to include a space-time dilation as:
\bea
\Y(x)&\to& \Y'(x') = \Y(x) + \Lambda c,   \label{eqn:sim-transformation-1} \\
\X(x)&\to& \X'(x') = e^{-c \Lambda/R_0}\X(x),    \label{eqn:sim-transformation-2} \\
x^\mu &\to& x^{\prime\mu} = e^{c} x^\mu . \label{eqn:sim-transformation-3}
\eea
Under this transformation, the action picks up an overall factor
\begin{equation}
 S \to S^\prime = e^{2c} S.
 \label{eqn:scaling-action}
\end{equation}
It immediately follows that under the previous transformations the classical equations of motion remain unchanged. For the same reasons, although this set of transformations do not constitute a true symmetry transformation, for the sake of simplicity we will refer to it as a symmetry. As we shall soon see, that the inflationary attractor solution respects the same scaling transformation. This implies that we can map different background solutions onto each other. This is the key property that we exploit in this work. We study the implications of this scaling transformation on the behavior of perturbations around the inflationary background solution.  \\

\subsection{Implications for perturbations: existence of an ultra-light field}
\label{subsection-ultra-ligth-def}

Given a particular background solution $(\bar{\X}(t)$, $\bar{\Y}(t))$ we can map it to a family of solutions, parametrized by the quantity $c$ appearing in the scaling transformation of Eqs.~(\ref{eqn:sim-transformation-1})-(\ref{eqn:sim-transformation-3}). To be precise, we can define new background solutions $(\X_c(t)$, $\Y_c(t))$ out from $(\bar{\X}(t)$, $\bar{\Y}(t))$ as:
\begin{align}
&\Y_c(t) = \bar{\Y}(e^{-c}t) +\Lambda c, \label{gen-back-1}  \\
&\X_c(t) =  e^{-c\Lambda/R_0}\bar{\X}(e^{-c}t),  \label{gen-back-2}
\end{align}
Now, we would like to define fluctuations about the background inflationary trajectory $(\bar{\X}(t)$, $\bar{\Y}(t))$ that represent deviations from homogeneity. The previous relation suggests that a natural way to parametrize perturbations around the homogeneous background solution is given by
\begin{align}
&\Y(x) = \bar{\Y}\left(t+\pi(x)\right)+\F(x), \label{eqn:perturbedbackground-1}  \\
&\X(x) =   e^{-\F(x)/R_0}\bar{\X}\left(t+\pi(x)\right). \label{eqn:perturbedbackground-2}
\end{align}
The fluctuations $\pi$ and $\F$ correspond to the two scalar degrees of freedom of the system. As we shall see in Section \ref{sec:physicalperturbations}, we may trade one of them (say $\pi$) for a metric perturbation through a gauge choice. Notice that the perturbation $\F$ fluctuates along the direction generated by the parameter $c$.    \\

It is particularly important to notice that configurations with constant values of $\F$ and $\pi$ correspond to an allowed background solution, which is evident from (\ref{gen-back-1}) and (\ref{gen-back-2}). This in turn implies that constant $\F$ and $\pi$ are allowed solutions of the perturbed system (a statement that we shall confirm in the next sections). Given that in an expanding FRW background the physical wavelength is stretched, we anticipate that the inhomogeneous equations of motion must admit a solution for $\F$ which becomes constant in the long wavelength limit.
In other words, $\F$ must freeze after horizon crossing. This means that we can identify $\F$ as an ultra-light field.  \\

\section{Background attractor solutions}
\label{sec:attractors}
The homogeneous background equations of motion describing the dynamics of the scalar fields evolving in a flat FLRW spacetime $
ds^2 = -dt^2 +a^2 d{\bf x}^2$, are given by
\begin{align}
&\ddot X + 3 H \dot X + \frac{2}{R_0} \dot Y \dot X + e^{- 2 Y / R_0  }V_X = 0 , \label{eqn:backgr1} \\
&\ddot Y + 3 H \dot Y - \frac{1}{R_0} e^{ 2 Y / R_0  } \dot X^2 +V_Y = 0 , \label{eqn:backgr2} \\
& 3 H^2  =  \frac{1}{2} e^{2 Y / R_0  } \dot X^2 + \frac{1}{2} \dot Y^2 +  V , \label{eqn:backgr3}
\end{align}
where $~\dot \  = d/dt$,  $H = \dot a / a$ is the Hubble expansion rate. One may combine these three equations to obtain a relation for $\dot H$, found to be given by:
\be
- \dot H = \frac{1}{2} e^{2 Y / R_0  } \dot X^2 + \frac{1}{2} \dot Y^2 \label{eqn:backgr4}   .
\ee
It is straightforward to show that the previous equations of motion are invariant under the scaling transformations (\ref{eqn:sim-transformation-1})-(\ref{eqn:sim-transformation-3}). Nevertheless, a generic solution to the equations of motion spontaneously breaks the symmetry.  Indeed, we may use the property of Eq.~(\ref{eqn:potentialidentity}) to integrate once the equations of motion~(\ref{eqn:backgr1})-(\ref{eqn:backgr3}). One finds the following relation valid for any background trajectory
\begin{equation}
C(t) \equiv e^{2Y/R_0} X\dot{X}-R_0\dot{Y} +2 \beta H =
\frac{C_0}{a^3}\ ,
\label{eqn:genericsolutionbackground}
\end{equation}
where $C_0$ is an integration constant set by the initial conditions of the background trajectory. It is straightforward to verify that $C(t)$ is not invariant under (\ref{eqn:sim-transformation-1})-(\ref{eqn:sim-transformation-3}), because the right-hand side of~(\ref{eqn:genericsolutionbackground}) scales differently than the left-hand side  (this can been seen most easily by absorbing the transformation of the spatial coordinates in a transformation of the scale factor $a \to a' = e^{c} a$). However, the system approaches the attractor regime $C = 0$ exponentially fast (in $e$-folds), at which point the symmetry is dynamically restored. We conclude that the inflationary background solution spontaneously breaks the symmetry, but it quickly gets \textit{dynamically restored} as $C\rightarrow 0$. The equation $C = 0$ can be used to replace one of the three equations~(\ref{eqn:backgr1})-(\ref{eqn:backgr3}), simplifying the search of solutions to the system.    \\

For later convenience, here we define the slow-roll parameters as
\be
\epsilon \equiv -\frac{\dot H}{H^2} , \qquad  \eta \equiv \frac{\dot \epsilon}{H \epsilon}.
\ee
For the upcoming analysis of perturbations, we also find it useful to define a dimensionless parameter $\Delta$ by:
\bea \label{Delta-intro}
1+ \Delta \equiv \frac{1-{\beta \dot Y}/{(\epsilon R_0 H)}}{\sqrt{1-{\dot Y^2}/{(2\epsilon H^2)}}}  .
\eea
The equations of motion allow one to show that $1+ \Delta$ can also be expressed as
\be
1+ \Delta = \sqrt{1+ \frac{X^2}{R_0^2}e^{2Y/R_0} -\frac{2\beta^2}{\epsilon R_0^2}} . \label{alt-Delta}
\ee
In Section~\ref{sec:squeezedlimitbispectrum} we will show that this background quantity relates the ultra-light field $\F$ (introduced in Section~\ref{subsection-ultra-ligth-def}) and the isocurvature mode $\sigma$ [introduced in Eq.~(\ref{action-quadratic-intro})], at linear level, as
\be
\sigma = (1 + \Delta) \F  \label{sigma-F-linear-relation} \  + \  O( \F^2) \ .
\ee
Additional relations among background parameters involving $\Delta$ can be derived by using the background equations of motion. We list these relations in Appendix~\ref{app:algebraicrelations}. \\

For the rest of this section we consider two main classes of attractor trajectories.
We shall show that the scaling transformation approach provides a possible unification for different realizations of  the ``ultra-light isocurvature" scenario. \\

In what follows we will use the notation
 $$s \equiv X  e^{Y/R_0},$$
 so the potential eq. (\ref{potential}) takes the form $V = X^{2\beta} G(s)$, with $G$ an arbitrary function.  \\

\subsection{Class 1: Symmetry probing attractors} \label{class1-SPS}

The existence of the symmetry ensures a class of inflationary trajectory that lies along the direction where scaling transformation is realized. It is straightforward to find that, in terms of $e$-folds $N$ (related to cosmic time $t$ through $d N = H dt$), the solutions are
\bea
X &=& X_0 e^{- Y_0' N / R_0} , \\
Y &=& Y_0 + Y_0' N ,
\eea
where $X_0$, $Y_0$ and $Y_0'$ are constants. The equations of motion require that $H$ evolves in time as
\be
H = H_0 e^{-\epsilon_0 N}, \qquad \epsilon = \epsilon_0
\ee
where $H_0$ and $\epsilon_0$ are additional constants.. Notice that in this class of backgrounds $s$ is constant:
\be
s \equiv X  e^{Y/R_0} = \rm{const} = \gamma  .
\ee
It turns out that the equations of motion imply conditions on all the constants except for one, which remains as a free parameter. Let us choose this parameter to be $X_0$. Then one finds that $\gamma$ must be a solution of the following algebraic equation
\be \label{gamma-algebraic-eq}
 \frac{4 \gamma^2 \beta^2}{3 (R_0^2 + \gamma^2) - 2 \beta^2} - 2 \beta R_0^2 - (R_0^2 + \gamma^2) \gamma \frac{G' (\gamma) }{G(\gamma)} = 0 .
\ee
When this equation is not satisfied for a real value of $\gamma$, this class of probing symmetry solution is not possible. On the other hand, we have verified numerically that, when the solution is stable, for arbitrary initial conditions the system quickly evolves towards a state in which $\gamma$ satisfies (\ref{gamma-algebraic-eq}).
Then, every background quantity can be determined in terms of $\gamma$ and $X_0$. Concretely, one finds
\bea
Y_0' &=& \frac{2 \beta R_0}{R_0^2 + \gamma^2} , \\
\epsilon_0 &=& \frac{2 \beta^2}{R_0^2 + \gamma^2 } , \\
H_0^2 &=& \frac{R_0^2 + \gamma^2 }{3R_0^2 + 3\gamma^2 - 2 \beta^2} X_0^{2 \beta}   G (\gamma).
\eea
Out of these quantities, one can additionally compute expressions for the aforementioned parameter $\Delta$, together with the rate of turn $\Omega$ and the entropy mass $\mu$. These are found to be given by
\bea
\Delta &=& - 1, \\
\frac{\Omega^2}{H^2} &=& \frac{4\beta^2 \gamma^2}{R_0^2(R_0^2+\gamma^2)^2} ,   \label{SPS-omega}  \\
\frac{\mu^2}{H^2} &=&  (3-\epsilon_0)\bigg( \frac{2 \beta^2}{R_0^2 \epsilon_0} \frac{G''}{G} + \left[\frac{4\beta}{\gamma} + \frac{\gamma}{R_0^2}\right] \frac{G'}{G} \nn \\
&& + \epsilon_0 \frac{(2 \beta-1) R_0^2}{\beta\gamma^2}  - \frac{2 \epsilon_0}{\beta} \bigg) - \frac{2 \epsilon_0}{R_0^2} + 3 \frac{\Omega^2}{H^2} .  \label{SPS-mu}
\eea
Recall from Eq.~(\ref{sigma-F-linear-relation}) that, to linear order, $\sigma = (1 + \Delta) \F$. It follows that in this class of trajectories the relation between $\sigma$ and $\F$ becomes ill defined. This result does not imply that $\sigma$ cannot be defined. It simply informs us that the fluctuation $\F$ does not constitute a good parametrization of the perturbed system, and so it cannot be identified with the isocurvature field. In fact, in the case $1 +\Delta =0$ the fluctuation $\F$ becomes directly related to the curvature perturbation $\R$, which indeed parametrizes fluctuations along the trajectory. We will come back to this issue when we study more closely the dynamics of fluctuations in Section~\ref{sec:physicalperturbations}.    \\

\subsection{Class 2: Misaligned  attractors} \label{class-misal}

The system under study also admits  attractor solutions along  trajectories misaligned with the symmetry-probing direction.
If the potential is appropriately chosen --consistent with eq. (\ref{eqn:potentialidentity})--, these solutions can satisfy the slow roll condition $\epsilon \ll 1$ during the whole period of inflation. We find that these solutions are characterised by $\Delta =$constant (but $\Delta \neq - 1$). There are two particularly interesting families of attractor trajectories within this class of solutions that we find worth discussing, which we do in what follows.   \\

\subsubsection{Trajectories with $\dot Y = 0$}
\label{sec:isometry}

It is interesting to notice that the present system includes potentials admitting trajectories such that $\dot Y = 0$. The most readily available example is given by the choice $G (s) = w_0^2( 3 -2 \beta^2 / s^2 )$. In this case, the potential can be written as
\be
V(X,Y) = w_0^2 X^{2 \beta} \left( 3 - 2  \frac{\beta^2}{X^2 e^{2 Y /R_0}}  \right) . \label{pot-JH-reduction}
\ee
To verify that this potential admits $\dot Y = 0$ solutions, it is enough to recognise that this potential can be written in terms of a ``fake" superpotential $W(X)$ as
\be
V(X,Y) = 3 W^2 - e^{-2 Y / R_2} W_X^2 ,
\ee
where $W(X) = w_0 X^\beta$ (in the previous expression, $W_X$ denotes a derivative of $W$ with respect to $X$). It turns out that this system automatically satisfies the following Hamilton-Jacobi equations:
\be
\dot X =  - \frac{1}{2} W_X , \qquad
\dot Y = - \frac{1}{2} W_Y, \qquad
H = W .
\ee
Given that in this example $W$ is independent of $Y$, it follows that the system has an attractor characterised by $\dot Y=0$. The present system with a potential given by~(\ref{pot-JH-reduction}) corresponds to {\it shift-symmetric orbital inflation} introduced in Ref.~\cite{Achucarro:2019pux} in the specific case where the field space has a hyperbolic geometry. We can in fact generalise the potential $V(X,Y)$ of Eq.~\eqref{potential} away from the specific Hamilton-Jacobi structure. In order to ensure $\dot Y = 0$, a potential of the form given in Eq.~\eqref{potential}  must agree with the following equations of motion
\begin{align}
&\ddot X + 3 H \dot X + e^{- 2 Y / R_0  }V_X = 0 , \label{eqn:backgr1-iso} \\
& V_Y =  \frac{1}{R_0} e^{ 2 Y / R_0  } \dot X^2 , \label{eqn:backgr2-iso} \\
& 3 H^2  =  \frac{1}{2} e^{2 Y / R_0  } \dot X^2  +  V. \label{eqn:backgr3-iso}
\end{align}
A potential satisfying these equations of motion must be such that its function $G(s)$ satisfies the following differential equation:
\be
s \frac{d}{ds}F+ \left( 2+ F \right) \left( 2 \beta+ F \right) = s \sqrt{6 F (2 + F)} , \label{eq-for-G}
\ee
where we have defined $F(s) \equiv s  \frac{d}{ds} \ln G$. It can be verified that the choice $G (s) = w_0^2( 3 -2 \beta^2 / s^2 )$ introduced earlier satisfies this equation, in which case $w_0$ turns out to be an integration constant. More general forms of $G (s)$ can be obtained from (\ref{eq-for-G}), a task that we do not examine in the present article.  \\

Now, from Eq.~(\ref{Delta-intro}) we see that this type of trajectory implies that $\Delta = 0$, and so $\F$ and $\sigma$ becomes exactly the same field. As we shall see soon, $\Delta = 0$ implies exactly that the entropy mass $\mu$ is zero, as expected. \\

\subsubsection{Trajectories with $\dot Y \neq 0$}
\label{sec:VY0}

It is also possible to have other trajectories that remain misaligned with respect to the symmetry probing solution, but without $\dot Y = 0$. Analytic results for this category are in general rather hard, but we can at least obtain reliable approximations in the particular case in which we can neglect $V_Y$ in the equation of motion~(\ref{eqn:backgr2}).
A concrete example was discussed in Ref. \cite{Achucarro:2016fby}, where  $V_Y = 0$ was taken to be exact. By neglecting $V_Y$ in Eq.~(\ref{eqn:backgr2}) and disregarding slow-roll corrections, the background equations of motion become
\begin{align}
&  3H\dot X  + e^{- 2 Y / R_0  }{V_X} = 0 ,  \\
& 3 H \dot Y= \frac{1}{R_0} e^{ 2 Y / R_0  }\dot X^2  .
\end{align}
These equations show that the slow-roll motion of $X$ is controlled by the slope of the potential, whereas $Y$ rolls slowly thanks to the centrifugal force imposed by the turning of the trajectory. The corresponding trajectories are given by
\bea
&& \dot X \simeq -\frac{2\beta H}{X} e^{-2Y/R_0} , \\
&& \dot Y \simeq \frac{4\beta^2H}{3R_0X^2} e^{-2Y/R_0} . \label{Ydot}
\eea
Notice that,  in this solution, both fields have non-vanishing velocity. The ratio of their physical velocities is given by
\be
e^{2Y/R_0} \dot X^2 /\dot Y^2 = \frac{9R_0^2}{4\beta^2}X^2e^{2Y/R_0} = \frac{9R_0^2}{2\epsilon_X} ,
\ee
where $\epsilon_X \equiv\frac{1}{2H^2} e^{2Y/R_0} \dot X^2\ll1$.
Thus for a not too small $R_0$, the kinetic energy of the $X$ field is much larger than the one of the $Y$ field.
As a result, the leading order contribution to $\epsilon$ is given by
\bea
\epsilon &\simeq&\epsilon_X = \frac{2\beta^2}{X^2} e^{-2Y/R_0} .
\eea
Using $\epsilon_X\ll1$ and expanding the background equations order by order, we can also compute $\epsilon$ up to next-to-leading order
\be \label{epsNTL}
\epsilon = \frac{2\beta^2}{X^2} e^{-2Y/R_0} \left( 1-\frac{2}{3\beta}\epsilon_X + \frac{2}{9R_0^2}\epsilon_X \right).
\ee
Next we compute the parameter $\Delta$ defined in \eqref{Delta-intro}. Interestingly for this class of trajectories, the leading term in $\epsilon$ is exactly cancelled in the expression of $\Delta$. Then the next-to-leading order \eqref{epsNTL} gives us
\be
(1+\Delta)^2 = \left( 1- \frac{2\beta}{3R_0^2}\right)^2 ,
\ee
where $\Delta$ is constant but non-zero. Notice that for $\beta>1$ and ${2\beta}\lesssim{3R_0^2}$, the system becomes part of the symmetry probing class for which $\Delta=-1$, described in Sec. \ref{class1-SPS}. This is a special case in the derivation above, which should not be seen as misaligned trajectories.   \\

We have examined numerically other varieties of attractor solutions with $\dot Y \neq 0$, away from the simpler case whereby $V_Y$ can be neglected, and have consistently found that as long as slow roll is imposed, they contain backgrounds with constant $\Delta$. Given that $\sigma = (1 + \Delta) \F$, it follows that, again, $\F$ and $\sigma$ can be identified, up to a proportionality constant. This will nevertheless still imply that the properties of $\F$ are inherited by $\sigma$ and, in particular, $\sigma$ will be ultralight for this type of background trajectories. We examine these statements more closely in the next sections.   \\

\section{Perturbations}\label{sec:physicalperturbations}

We now derive the quadratic action for the perturbations defined in Section~\ref{sec:dualitytransformation}. This derivation will allow us to understand the role of the scaling symmetry, and will confirm our intuition about how it ensures the existence of an ultra-light field. The perturbed fields are given in (\ref{eqn:perturbedbackground-1})-(\ref{eqn:perturbedbackground-2}), where we write the perturbed metric using a slightly modified ADM-like decomposition
\begin{equation}
 ds^2 = e^{2\F}\Big( - dt^2   N^2 + a^2 e^{2 \tilde\varphi} \delta_{i j} (dx^i + N^i dt)(dx^j + N^j dt) \Big), \label{metric-pert}
\end{equation}
where $N$ and $N^i$ are the lapse and shift functions. Notice that $\F$ appears as a dilatonic fluctuation multiplying the entire metric. In order to have the usual definition of the lapse, we should actually absorb the factor $e^{2\F/R_0}$ in $N$. Nevertheless, in what follows we will keep the definition of $N$ as in (\ref{metric-pert}). Finally, we pick a new gauge in which $\pi(x)=0$.
In this gauge the metric fluctuation $\tilde\varphi$, representing spatial curvature perturbations, becomes physical. However, notice that this is \textit{not} the comoving gauge, since the perturbations along the trajectory (which involves $\F$) are non-vanishing. We call it the {\it ultra-light gauge}, and will establish its connection to the comoving gauge in Section~\ref{sec:connection-two-gauges} and Appendix \ref{sec:gauge}. By inserting the perturbed metric (\ref{metric-pert}) and the perturbed fields of (\ref{eqn:perturbedbackground-1})-(\ref{eqn:perturbedbackground-2}) back into the original action (\ref{basic_action}), we find that the full perturbed action (to all orders) is given by
\begin{widetext}
 \begin{eqnarray}
S &=& \frac{1}{2} \int d^4 x a^3 e^{3 \tilde\varphi} e^{2n\F/R_0}N  \bigg[ - \frac{2}{a^2} e^{- 2 \tilde\varphi} \left[ 2 \nabla^2 (\tilde\varphi+n\F/R_0) + (\nabla (\tilde\varphi+n\F/R_0))^2 \right] \nn \\
&&
 + \frac{1}{2 N^2} \left( N^{i}{}_{,j}N^{j}{}_{,i}  + \delta_{ij} N^{i,k} N^{j}{}_{,k} - 2 N^{i}{}_{,i} N^{j}{}_{,j} \right) - \frac{6}{N^2} \left( H + \dot {\tilde\varphi} + n\frac{\dot{\F}}{R_0} - N^i \tilde\varphi_{, i} -nN^i \frac{{\F}_{,i}}{R_0}\right)^2  \nn \\
 && + \frac{4}{N^2} \left( H + \dot {\tilde\varphi} + n\frac{\dot{\F}}{R_0} - N^i \tilde\varphi_{, i} -nN^i \frac{{\F}_{,i}}{R_0} \right) N^{j}_{,j}   + \frac{1}{N^2} e^{ 2 \bar Y / R_0 } \left(\dot {\bar X}  - \bar X   \frac{\dot \F}{R_0} +  N^i \bar X   \frac{\F_{,i}}{R_0}\right)^2  \nn \\
&&
 + \frac{1}{N^2} (\dot{\bar Y}+\dot \F - N^i \F_{, i})^2 -  e^{ 2 \bar Y / R_0 } \frac{\bar X^2}{R_0^2} h^{ij} \F_{, i} \F_{, j} - h^{ij} \F_{, i} \F_{, j}  - 2V(\bar X, \bar Y)  \bigg]   .  \label{action_pert}
\end{eqnarray}
\end{widetext}
We see that the only place where $\F$ appears without space-time derivatives acting on it, is in the overall exponential factor. This means that its equations of motion will be equal to a functional of derivatives of $\F$ and $\tilde\varphi$, and the lapse and shift. Therefore, as long as the lapse and shift don't depend on $\F$ without space-times derivatives acting on it, it follows that $\F$ admits a constant solution in the long wavelength limit.  \\

To obtain the quadratic (or cubic) action for the perturbations, we need first order solutions of the lapse and shift functions. For notational convenience we define
\begin{equation}
 \varphi \equiv \tilde\varphi+\beta \F/R_0\ .
\end{equation}
Varying the perturbed action (\ref{action_pert}) with respect to $N^i$, one finds the following constraint on the lapse $N$:
\be
N = 1 + \frac{\dot{\varphi}}{H} -\frac{C(t)}{2H}\frac{\F}{R_0} , \label{N-constraint}
\ee
where $C(t)$ is the background quantity defined in Eq.~\eqref{eqn:genericsolutionbackground}.
On the other hand, varying the action with respect to $N$, one finds that the shift function must satisfy
\begin{equation}
 \partial_i N^i = -\frac{\nabla^2\varphi}{a^2 H} -(3-\epsilon)H \delta N + 3\dot\varphi +\left(\frac{C(t)}{2H}-\beta \right)\frac{\dot\F}{R_0}\ .
\end{equation}
Whenever $\delta N$ is independent of $\F$, it follows that $\partial_i N^i$ will be independent of $\F$ as well.
Notice that the dependence of $N$ on $\F$ is proportional to $C(t)$. Its presence breaks the shift symmetry in $\F$. This makes sense, because the non-attractor solution does not respect the scaling transformation, as we saw in \autoref{sec:dualitytransformation}. The symmetry is restored as soon as the system converges to the attractor $C = 0$. Therefore, we have explicitly shown that to third order in perturbation theory $\F$ indeed admits a constant solution.   \\

Let us now deduce the quadratic action for the perturbations on the attractors. By inserting (\ref{N-constraint}) back into (\ref{action_pert}) and expanding up to second order, we finally deduce
 \begin{eqnarray}
S &=& \frac{1}{2} \int d^4 x a^3  \bigg[ -\frac{2\epsilon}{a^2} (\nabla \varphi)^2 +\frac{4\beta}{a^2R_0}\nabla \F \cdot \nabla\varphi  \nn \\
&& - \left(1+ \frac{\bar X^2}{R_0^2}e^{ 2 \bar Y / R_0 }  \right)\frac{(\nabla \F)^2}{a^2} + 2\epsilon \left(\dot \varphi-\frac{\beta \dot\F}{\epsilon R_0}\right)^2 \nn \\
 &&  +\left(1+\frac{\bar X^2}{R_0^2}e^{ 2 \bar Y / R_0 }-\frac{2\beta^2}{R_0^2\epsilon}\right)\dot\F^2 \bigg], \label{quadratic-action}
 \end{eqnarray}
where we have assumed that $C(t) \to 0$ has settled. Thanks to Eq.~(\ref{alt-Delta}), we see that $\F$ is well defined as long as $\Delta \neq -1$, which excludes the first class of trajectory solutions studied in Section~\ref{class1-SPS}. In the case $\Delta = -1$, we must therefore directly work with the action (\ref{action-quadratic-intro}), with parameters given in Eqs.~\eqref{SPS-omega} and \eqref{SPS-mu}.  \\

Notice that~\eqref{quadratic-action} is symmetric under constant shifts of $\F$, confirming the reasonings of \autoref{sec:dualitytransformation}, based on the invariance of the classical equations of motion. A simple inspection of the equations of motion for $\varphi$ and $\F$ shows that $\left(\dot \varphi-\frac{\beta \dot\F}{\epsilon R_0}\right)$ quickly decays to zero during inflation. Therefore, we can infer from the $\F$ equation of motion for that $\F=\text{const}$ is the dominant solution. The other solution is decaying on superhorizon scales. This allows us to trade a long wavelength $\F$ perturbation for a scaling transformation of the background solution.  \\

\section{Consequences of the scaling transformation}\label{sec:scalingtransformation}

In this section, we study a few direct consequences of the scaling transformation on the behavior of the field $\F$. For instance, we may infer the squeezed limit of the 3-point function of $\F$ from the way in which the scaling transformation acts on $\F$ and the background quantities. We will use this result in the next section to derive the squeezed limit of the bispectrum of the primordial curvature perturbation $\R$. \\

\subsection{Scaling transformation and perturbations}  \label{sec:sim-perturbations}

Let us revisit how the scaling transformations (\ref{eqn:sim-transformation-1})-(\ref{eqn:sim-transformation-3}) act on the perturbed fields $\X(x)$ and $\Y(x)$ given in Eqs.~(\ref{eqn:perturbedbackground-1})-(\ref{eqn:perturbedbackground-2}). Notice that the scaling transformation can be re-stated as a transformation on the fluctuation $\F(x)$ and the coordinates $x$, instead of on the fields $X$ and $Y$ (which only appear as background quantities). That is, we may summarize the transformation as:
\begin{eqnarray}
\mathcal F  (x)  =  \mathcal F ' (x')  + c , \qquad x' = e^{c} x . \label{transf-F}
\end{eqnarray}
(Recall that we are working in a gauge where $\pi = 0$). Now the background fields $X$ and $Y$ transform as scalars:
\begin{align}
&X(t) \to X'(t') = X (t) , \label{X-t'-t} \\
&Y(t) \to Y'(t') = Y(t) .  \label{Y-t'-t}
\end{align}
That is, the only effect of the transformation is to rescale the time coordinate, without affecting the field. If needed, now one could rewrite (\ref{X-t'-t}) and (\ref{Y-t'-t}) as $X'(t) = X(e^{-c}t)$ and $Y'(t) = Y(e^{-c}t)$. Equations~(\ref{transf-F})-(\ref{Y-t'-t}) imply that any background quantity will transform according to how they scale with time. For example, it is straightforward to see that the scale factor also transforms as a scalar $a(t) \to a'(t') = a(t)$, and therefore the Hubble parameter transform as:
\be
H(t) \to H'(t') = e^{-c} H(t) .
\ee
We can now identify $\F'(x')$ as a fluctuation about a background determined by $X'(t')$, $Y(t')$ and $H'(t')$, that differs from the original background by a re-scaling of time. However, because the transformation is not a symmetry of the action, there is an important distinction between $\F$ and $\F'$. Indeed, the transformations~(\ref{transf-F})-(\ref{Y-t'-t}) imply that the action~(\ref{action_pert}) for the fluctuations rescales as $S \to S' = e^{2c} S$, as expected. This re-scaling tells us that the quantization of $\F$ and $\F'$ follow slightly different rules. To be precise, while $\F$ is quantized employing $\hbar$, the fluctuation $\F'$ must be quantized using a value of the Planck constant given by
\be
\hbar' = e^{2c} \hbar,
\ee
which is obtained by requiring that $S' / \hbar ' = S / \hbar$, after considering $S \to S' = e^{2c} S$.  \\

\subsection{Power spectrum of $\F$}

Let us use the previous results to establish a simple consequence concerning the computation of the 2-point function for $\F$ that will be useful in the next subsection. First, recall that the quadratic action~(\ref{quadratic-action}) is symmetric under shifts of the field $\F$ (without considering the simultaneous re-scaling the coordinates). This implies that the $\F$ field must admit as a solution a constant amplitude. In particular, at long wavelengths (after horizon crossing) $\F$ will freeze to a constant value (because the other non-constant solutions turn out to be decaying solutions). To make this statement more tractable, let us expand $\F$ in Fourier modes as
\be
\mathcal F ( {\bf x} ) = \frac{1}{(2 \pi)^3} \int d^3 k \hat \F_{\bf k} (t) e^{ i {\bf k} \cdot {\bf x}} .
\ee
At linear order, $\F_{\bf k} (t)$ may be expanded in terms of creation and annihilation operators as:
\be
\hat \F_{\bf k} (t) = \F_k (t) a_{\k} + \F_k^* (t) a^{\dag}_{-\k} .
\ee
The statement that $\F$ freezes on superhorizon scales is equivalent to say that the amplitude $\F_k (t)$ becomes a constant at wavelengths such that $k / a (t) \ll H(t)$. The value of $\F_k (t)$ on superhorizon scales must be given by a function of background quantities. This is because the coefficients of the action~(\ref{quadratic-action}) depend only on background quantities such as $a(t)$, $\epsilon$, and combinations $\X^2 e^{2 \Y / R_0}$. Then, from dimensional analysis, at superhorizon scales $\F_k (t)$ must be proportional to $k^{-3/2}$ times a given constant $B$ with dimensions of $t^{-1}$ that is determined by background quantities (for instance, $B$ could be given by $H_*$, where the $*$ denotes that $H$ is evaluated at the time of horizon crossing). In other words:
\be
\F_k (t)  \to k^{-3/2} B .
\ee
The dimensionless power spectrum $\mathcal P_{\F} (k)$ may be defined as the amplitude of the $2$-point function as:
\be
\langle \mathcal F \mathcal F \rangle ( \k_1 , \k_2)  = (2 \pi)^3 \delta^{(3)} (\k_1 + \k_2) \frac{2 \pi^2 \P_{\F} }{k^3} .
\ee
Then, it follows that
\be
\P_{\F} = \frac{B^2 \hbar }{ 2 \pi^2} .
\ee
This looks like a scale invariant power spectrum. However, $B$ will necessarily depend on $k$ through its dependence on the background quantities. More to the point, $B$ is a constant function of background quantities that are evaluated at a time $t_*$, determined by the equation $H(t_*) a(t_*) = k$ (which signals horizon crossing). This implies that $B$ inherits a $k$ dependence, that will be small as long as the background quantities on which it depends are slowly evolving in time (as expected if $\epsilon$ and $\eta$ are small). \\

Now, if we had computed the power spectrum of $\F'$ instead of $\F$, then we would have considered a background $B '$ and a value of the Planck constant given by $\hbar'$. This would have led to the result
\be
\P'_{\F}  = \frac{{B'}^2 \hbar' }{ 2 \pi^2} .
\ee
Then, because $B$ has dimensions of $t^{-1}$, we necessarily have $B '  =  e^{-c} B$. In addition, recall that $\hbar' = e^{2c} \hbar$. These two transformation rules then imply that
\be
\P'_{\F} =  \P_{\F} . \label{P'=P}
\ee
That is, the power spectrum remains invariant under the scaling transformation. \\

\subsection{The squeezed limit of the 3-point function}

Let us now show how to compute the squeezed limit of the 3-point function of $\F$. Here we follow the background-wave method introduced in~\cite{Maldacena:2002vr} to derive the squeezed limit (see also~\cite{Creminelli:2004yq, Cheung:2007sv, Bravo:2017wyw}). First, we may split $\mathcal F (x)$ into short and long-wavelength contributions as $\mathcal F = \mathcal F_S + \mathcal F_L$. This splitting implies that $\mathcal F_L (x)$ contains contributions from wavelengths that exited the horizon at times earlier than $\mathcal F_S (x)$. Given that $\mathcal F$ freezes at horizon crossing, and given that the gradients of $\mathcal F_L$ are more suppressed than those of $\mathcal F_S$, we can take $\mathcal F_L$ as a constant (in comparison to $\mathcal F_S$). Then, setting $c = \mathcal F_L$ in the transformation rule (\ref{transf-F}), we obtain:
\begin{eqnarray}
\mathcal F_S  (x)  =  \mathcal F ' (e^{\mathcal F_L} x)  . \label{F_S-F'}
\end{eqnarray}
Now, in the previous section we learned how to compute the power spectrum of  $\mathcal F ' (x')$. Let us write this result as
\be
\langle \mathcal F'  (x') \mathcal F'  (y') \rangle = \langle \mathcal F' \mathcal F' \rangle  (|{\bf x}' - {\bf y}' | ) .
\ee
As a consequence, relation (\ref{F_S-F'}) allows us to compute the power spectrum of $\mathcal F_S  (x)$ and express it as a function of $\mathcal F_L$:
\begin{eqnarray}
\langle \mathcal F_S  \mathcal F_S   \rangle   ( |{\bf x} - {\bf y} | )  = \langle \mathcal F' \mathcal F' \rangle  (e^{\mathcal F_L} |{\bf x} - {\bf y} | )  \rangle .
\end{eqnarray}
Now, from Eq.~(\ref{P'=P}) we see that the only dependence of the right hand side on $\F_L$ is through the argument. This implies that the previous expression can be simply rewritten as
\begin{eqnarray}
\langle \mathcal F_S  \mathcal F_S   \rangle   ( |{\bf x} - {\bf y} | )  = \langle \mathcal F \mathcal F \rangle  (e^{\mathcal F_L} |{\bf x} - {\bf y} | )  \rangle .
\end{eqnarray}
where $\langle \mathcal F \mathcal F \rangle (|{\bf x} - {\bf y} | )$ is nothing but the 2-point function of $\F$ (instead of $\F'$) using the linear theory. The salient point of the previous relation is that it informs us how does $\F_L$ non-linearly affects $\F_S$. We may now expand this result in terms of $\mathcal F_L$ as
\begin{eqnarray}
\langle \mathcal F_S  (x) \mathcal F_S  (y) \rangle &=& \langle \mathcal F \mathcal F \rangle  ( |{\bf x} - {\bf y} | )  \rangle \nonumber \\
&& + \mathcal F_L  \frac{d}{d \ln |{\bf x} - {\bf y}|}\langle \mathcal F \mathcal F \rangle  ( |{\bf x} - {\bf y} | ) .  \qquad
\end{eqnarray}
Next, by Fourier transforming the fields as
\be
\mathcal F ( {\bf x} ) = \frac{1}{(2 \pi)^3} \int d^3 k \mathcal F_{\bf k} e^{ i {\bf k} \cdot {\bf x}} ,
\ee
we obtain the following relation in momentum space
\bea
\langle \mathcal F_S \mathcal F_S \rangle ( \k_1 , \k_2) &=&  \langle \mathcal F \mathcal F \rangle ( \k_1 , \k_2) \nn \\
&&  - \F_L (\k_L) (n_{\F} - 1) P_{\F}  (k_s) ,
\eea
where $\k_s = (\k_1- \k_2) / 2$ and $\k_L = \k_1+ \k_2$, and where $P_{\F} (k_s)$ is the power spectrum satisfying $\langle \mathcal F \mathcal F \rangle ( \k_1 , \k_2)  = (2 \pi)^3 \delta^{(3)} (\k_1 + \k_2) P_{\F} (k_s)$. In addition, we have defined the spectral index of $P_{\F} $ as
\be
n_{\F}(k_s) - 1  = \frac{d}{d\ln k_s} \ln \left[ k_s^3  P_{\F} (k_s) \right] .
\ee
We can now correlate the 2-point correlation function $\langle \mathcal F_S \mathcal F_S \rangle$ with the long mode $\F_L (\k_3)$. This gives us an expression for the squeezed limit of the 3-point function (where $|\k_3|\ll |\k_1|,|\k_2|$), which is found as:
\bea
&&\langle \mathcal F_S (\k_1) \mathcal F_S (\k_2) \mathcal F_L (\k_3)  \rangle_{\rm sq} = \nn \\
&& (2 \pi)^3  \delta^{(3)}(\sum_i \k_i) (1-n_{\F}) P_{\F}  (k_L) P_{\F}  (k_s) .
\label{eqn:wardidentity}
\eea
Therefore the scaling transformation property of the ultra-light field also suppresses its  self-interaction, and thus the bispectrum of $\F$ becomes negligible.
This is the main result of this section, which will be used in the following section to derive the bispectrum of the primordial curvature perturbation $\R$.  \\

\section{Comoving gauge and Squeezed bispectrum}\label{sec:squeezedlimitbispectrum}

As we have already stressed, the ultra-light fields $\varphi$ and $\F$ are not the usual curvature and isocurvature perturbations (hereby denoted as $\R$ and $\sigma$) in the comoving gauge. Nevertheless, they can be related by a gauge transformation. We leave the detailed derivation for the change of variables between two gauges up to second order to Appendix \ref{sec:gauge}. In this section, based on our previous results in the ultra-light gauge, we explore the perturbation behavior in the comoving gauge, and compute the squeezed limit of the bispectrum for the primordial curvature perturbation. \\

\subsection{Back to comoving gauge} \label{sec:connection-two-gauges}

By comparing the quadratic actions (\ref{action-quadratic-intro}) and (\ref{quadratic-action}) it is straightforward to realize that, to linear order, the two sets of perturbations are related as (see Appendix \ref{sec:gauge}):
\bea
\R &=& \varphi - \frac{\beta}{\epsilon R_0} \F , \label{eqn:gaugeR} \\
\sigma &=& (1+ \Delta) \, \F , \label{eqn:gaugetransformationsigma}
\eea
where we identified $\Delta$ with the help of Eq.~(\ref{alt-Delta}). In terms of these new perturbation variables, the action (\ref{quadratic-action}) becomes the well known expression already reported at the introduction:
\bea
S &=& \int d^4 x a^3  \bigg[  \epsilon \left(\dot \R  - \frac{2 \Omega}{\sqrt{2 \epsilon}}   \sigma\right)^2 -\frac{\epsilon}{a^2} (\nabla \R)^2 \nn \\
 &&  + \frac{1}{2}  \dot \sigma^2 - \frac{1}{2 a^2} (\nabla \sigma)^2  - \frac{1}{2}  \mu^2 \sigma^2 \bigg] ,
\eea
where $\Omega$ is the turning rate of the inflationary trajectory in field space, and $\mu$ is the entropy mass of the isocurvature field. In terms of $\Delta$ and slow-roll parameters, they are found to be given as
\bea
\Omega &=& \frac{\beta \eta H }{ \sqrt{2 \epsilon}  (1+ \Delta)  R_0} , \label{eqn:turnrate}\\
\mu^2 &=&  -\frac{\ddot \Delta}{1+\Delta} - 3 H \frac{\dot \Delta}{1+\Delta}  . \label{eqn:entropymass}
\eea
These results, especially Eq.~\eqref{eqn:entropymass}, confirm an important point: that the isocurvature field $\sigma$ inherits the properties of the ultralight field $\F$, and becomes ultra-light itself, provided that $\Delta$ is a constant along the trajectory (which, as we have seen, it is ensured by the symmetry of the UV model) which, as we have seen, can be the consequence of the underlying symmetry. In addition, $\Omega$ is in general non-vanishing, and so the interaction between the isocurvature field $\sigma$ and the curvature perturbation $\R$ remains turned on throughout inflation.  \\

In what follows, we derive the squeezed limit of the bispectrum of curvature perturbations for the class of models in which $\Delta$ remains constant, but different from the value $-1$. We start with the simpler case $\Delta = 0$, and then move on to consider the more general case $\Delta \neq 0$.  \\

\subsection{Bispectrum in the case $\Delta = 0$}\label{subsec:bispectrumzerodelta}

Let us first look at the exact ultra-light case with $\Delta = 0$ (or equivalently, $\dot Y = 0$) studied in Section \ref{sec:isometry}. Notice from \eqref{eqn:gaugetransformationsigma} that the ultra-light field $\F$ can be identified with the isocurvature perturbation $\sigma$, due to the fact that $\Delta = 0$ in this case.
Therefore $\sigma$ remains constant on superhorizon scales, and as a canonically normalized scalar field, the frozen amplitude is given by $\sigma_* = \F_* =H/(2\pi)$.
We denote the amplitude of the comoving curvature perturbation around horizon crossing as
\be
\R_* = \varphi_* - \frac{\beta}{\epsilon_*R_0} {\F_*} ,
\ee
whose amplitude is of order $H_*/\sqrt{\epsilon_*}$.
At the end of inflation, we have $\epsilon = 1$, and since the ultra-light fields $\F$ and $\varphi$ remain constant, the final amplitude of $\R$ can be expressed as
\be
\R(t_{\rm end}) = \varphi - \frac{\beta}{{R_0}} {\F} \simeq \R_* + \frac{\beta}{\epsilon_*R_0} {\F_*} .
\ee
Comparing these two terms, we find the dominant contribution at the end of inflation comes from the second one, which is of order $H_*/\epsilon_*$. Therefore the final curvature perturbation can be seen as mainly generated by the isocurvature degree of freedom, which is the essence of the ``ultra-light isocurvature" scenario \cite{Achucarro:2016fby, Achucarro:2019pux}.
More specifically, this requires $\epsilon_* R_0^2 /\beta^2 \ll 1$.
Then the power spectrum is given by
\be \label{powerR}
P_\R \simeq \frac{\beta^2}{\epsilon_*^2 R_0^2} P_\F .
\ee

We now derive the soft limit of the bispectrum of the curvature perturbation.
To compute the correlation functions of the comoving curvature perturbation, first we perform the gauge transformation from the ultra-light gauge to the comoving gauge to second order in perturbation theory (see Appendix \ref{sec:gauge})
\begin{align}
 \R = & ~\varphi - \frac{\beta}{\epsilon} \frac{\F}{R_0} -\left(\frac{\eta \beta^2}{4\epsilon^2}+\frac{\beta}{2\epsilon}\right)\frac{\F^2}{R_0^2} \nonumber \\
 & +\frac{\beta^2}{\epsilon}\frac{\dot \F \F}{R_0^2 H} - \frac{\beta}{\epsilon} \frac{\dot\varphi \F}{H R_0} .
\end{align}
Since $\varphi$ and $\F$ freeze out after horizon-exit, the last two terms will go to zero.
Then up to second order, the curvature perturbation at the end of inflation can be expressed as
\be
\R(t_{\rm end}) \simeq \R_* + \frac{\beta}{\epsilon_*} \frac{\F_*}{R_0} + \left(\frac{\beta}{2\epsilon_*}+\frac{\eta_* \beta^2}{4\epsilon_*^2}\right)\frac{\F_*^2}{R_0^2},
\label{eqn:gaugetransformation}
\ee
where we assume that $\eta_\text{end} \ll \eta_*/\epsilon_*^2$.
We have now at hand all ingredients to derive the soft limit of the bispectrum of the curvature perturbations.  The relation \eqref{eqn:gaugetransformation} allows us to express the bispectrum of curvature perturbations in terms of the bispectrum $\langle\F\F\F\rangle$ and four convolutions of the trispectrum of $\F$ evaluated at horizon crossing. Taking the squeezed limit of the bispectrum, we use the relation \eqref{eqn:wardidentity} to rewrite the squeezed bispectrum of $\F$ as function of its power spectrum. Moreover, considering the ultra-light field are the dominant source of the final curvature perturbation, we find
\begin{widetext}
\begin{align}
\lim_{k_l/k_s\rightarrow 0}  \langle \R(k_l)\R(k_s)\R(k_s) \rangle &
= (1-n_\F)\left(\frac{\beta}{\epsilon_\ast R_0}\right)^3 P_\F(k_l)P_\F(k_s) +\left(\frac{\beta}{\epsilon_\ast R_0}\right)^4\left(\frac{2\epsilon_\ast}{\beta}+ \eta_*\right) P_\F(k_l)P_\F(k_s) \nonumber \\
& \approx \left(\frac{2\epsilon_\ast}{\beta}+ \eta_*\right) P_\R(k_l)P_\R(k_s)~.
\end{align}
\end{widetext}
As expected, the tree-level contribution from the isocurvature self-interaction is suppressed with respect to the curvature bispectrum produced at super-horizon scales. Therefore, the final amplitude is given by
\begin{equation}
 f_{\rm NL}=\frac{5}{12}\left(\frac{2\epsilon_\ast}{\beta}+ \eta_*\right) .
 \label{eqn:squeezedlimitbispectrum1}
\end{equation}
This expression is consistent with the bispectrum of the orbital inflationary models with power-law potentials studied in \cite{Achucarro:2019pux}, for which $\eta= 2\epsilon/\beta$, applied to the hyperbolic field metric \eqref{eqn:K}. The resulting amplitude is slow-roll suppressed, however not equivalent to the single field inflationary prediction, because it violates Maldacena's consistency relation \cite{Maldacena:2002vr} $f_{\rm NL}=\frac{5}{12}\left(1-n_s\right) = \frac{5}{12}\left(2\epsilon_*+2\eta_*\right)$.  \\

\subsection{Bispectrum in the case with constant non-vanishing $\Delta$}\label{subsec:bispectrumconstantdelta}

Next, we focus on those background trajectories where $V_Y$ can be neglected, such that $\Delta$ is a non-vanishing constant, as discussed in Section \ref{sec:VY0}. We have verified numerically that the results reported in this section are re-obtained in other situations where $V_Y$ cannot be neglected. Here, because $\sigma$ and $\R$ are proportional to each other, the isocurvature field $\sigma$ freezes to an amplitude determined by $\sigma_* = (1+\Delta) \F_* =H_*/(2\pi)$. Meanwhile, since the linear order gauge transformation to $\R$ is given by \eqref{eqn:gaugeR}, the final amplitude of curvature  perturbation remains the same with \eqref{powerR}. For the squeezed limit of the bispectrum
we repeat the same steps as in Section~\ref{subsec:bispectrumzerodelta}, with the only difference that the non-linear gauge transformation changes. The full expression is derived in Appendix \ref{sec:gauge}. The final amplitude of the bispectrum is proportional to the coefficient of $\frac{\F^2}{R_0^2}$ of the gauge transformation and we find
the curvature perturbation at the end of inflation
\bea
\R(t_{\rm end}) &\simeq& \R_* + \frac{\beta}{\epsilon_*} \frac{\F_*}{R_0} +
\left[\frac{\eta}{4} + \frac{\epsilon}{2 \beta} -\frac{\dot Y^2}{2 \beta H^2}
\right.\nn\\
&& \left.
-\frac{2\dot Y}{R_0H}\left(1-\frac{\dot Y^2}{2\epsilon H^2}\right)+\frac{\epsilon R_0}{4\beta^2}\frac{\dot Y}{H} \right.\nn\\
&& \left.
-\frac{\epsilon R_0}{4\beta^2}\frac{\dot Y}{H}\left(4-\frac{\dot Y^2}{\epsilon H^2}\right)(1+\Delta)^2\right]
\frac{\beta^2\F_*^2}{\epsilon_*^2 R_0^2},
\eea
Next using the relations we derived in Section \ref{sec:VY0}, and keeping the leading order contribution, we find
\be
 f_{\rm NL}=\frac{5}{12}\left( \eta_* +\frac{2}{\beta}\epsilon_\ast  - \frac{16}{3R_0^2}\epsilon_* \right)\ ,
 \label{eqn:squeezedlimitbispectrum2}
\end{equation}
which is slow-roll suppressed {and violates Maldacena's consistency relation}. \\

\section{Discussion \& conclusions}
\label{sec:conclusions}

The present article was driven by the following questions: What is the active mechanism leading to the appearance of ultra-light isocurvature fields? And what is the generic prediction for non-Gaussianity from models admitting ultra-light isocurvature fields? We have seen that there are entire classes of multi-field systems that can sustain inflation with the very distinctive feature of having an ultra-light isocurvature fluctuation interacting with the primordial curvature perturbation. These systems are characterized by having an action that scales under a special class of transformation (a scaling transformation) that simultaneously acts on fields and space-time coordinates.  \\

This transformation leaves invariant the background equations of motion, and ensures the existence of two main classes of trajectories. The first class, introduced in Section~\ref{class1-SPS}, consists of trajectories that lie along the symmetry direction. In this case, the isocurvature field is not protected against acquiring a non-vanishing mass. In the second class, introduced in Section~\ref{class-misal}, the trajectory stays misaligned with respect to the first class, allowing the isocurvature field to fluctuate along the symmetry  direction
(when the parameter $\Delta$ defined in (\ref{Delta-intro}) stays constant). In such a case, the isocurvature field becomes ultra-light and it is possible to infer some outstanding properties about the primordial spectra. In particular, we were able to infer the squeezed limit of the bispectrum of the ultra-light fields, which is found to be suppressed by factors of order slow-roll.  \\

One of the main aspects of the present article that we wish to highlight is that the scaling property of the action ensures the existence of ultra-light fields for a long period of time during inflation ---at least long enough to affect the entire set of curvature perturbations relevant to observable scales. Understandably, the same mechanism ensuring the existence of ultra-light fields does not only suppress the value of the entropy mass, but also, it suppresses any other self-interaction experienced by the isocurvature field. As a consequence, even though ultra-light fields can efficiently transfer their statistics to the curvature perturbation, the amount of non-Gaussianity available for such a transfer happens to be marginal, and so the predicted level of primordial non-Gaussianity is found to be tiny. \\

All in all, one can now have a more general understanding on how primordial non-Gaussianity, as produced by an isocurvature field, can be enhanced. As already emphasized several times during this work, the existence of sizeable self-interactions is expected to come together with a non-vanishing entropy mass. Thus, after horizon crossing, there will be a limited amount of time during which the non-Gaussianity produced by these self-interactions can be transferred to the curvature perturbation. A clear example of this situation is offered by quasi-single field models of inflation~\cite{Chen:2009we, Chen:2009zp}, where a cubic self interaction of the isocurvature field is able to induce the appearance of local non-Gaussianity thanks to the linear coupling between the curvature and isocurvarture fields induced by $\Omega$. In this case, the amplitude of the bispectrum is found to be proportional to the cubic self-coupling of the isocurvature field, but it is also found to be suppressed by a given function of $\mu /H$. On the other hand, there are new examples of multi-field models where the self-interactions of the isocurvature field are determined by a scalar potential with a rich field structure, leading to the generation of tomographic non-Gaussianity~\cite{Chen:2018uul, Chen:2018brw} (whereby the non-Gaussian probability distribution function stores information of the shape of the landscape potential). In these cases, the structure of the potential ---which introduces a mass to the isocurvature field--- also introduces the self-interactions, leading to a controlled small enhancement of non-Gaussianity (in which high $n$-point functions could play a significant role). \\

Another important thing to notice is that, in the ``ultra-light isocurvature" scenario we studied, only one single degree of freedom (the ultra-light field) is responsible for both curvature and isocurvature perturbations by the end of inflation. This fact, besides the suppressed isocurvature self-interactions, also plays a crucial role in  the result of small non-Gaussianity. If we go beyond the ``ultra-light isocurvature" scenario, it is interesting to see similar behaviour also happens in ``multi-field $\alpha$-attractors" \cite{Achucarro:2017ing}, where although the multi-field effects are significant, in the end only the radial field fluctuation will contribute to the curvature perturbation, and its self-interaction is suppressed by the hyperbolic stretching effect. As a result, all the model predictions, including non-Gaussianity, recover the single-field results. These different studies of multi-field models indicate that inflation with unstabilized light fields can still yield single-field-like phenomenologies, and the origin of this result is related to the one, dominating single degree of freedom of perturbations with suppressed self-interaction. \\

\begin{acknowledgements}
We are grateful to Diederik Roest, Guilherme Pimentel and Rafael Bravo for comments and discussions. We thank Ed Copeland, Oksana Iarygina and Valeri Vardanyan for collaboration on related works.
The work of AA is partially supported by the Netherlands' Organization for Fundamental Research in Matter (FOM), by the Basque Government (IT-979-16) and by the Spanish Ministry MINECO (FPA2015-64041-C2-1P). GAP acknowledges support from the Fondecyt Regular project number 1171811 (CONICYT). GAP is grateful to the Lorentz Institute of Leiden University for hospitality.  DGW and YW are supported by a de Sitter Fellowship of the Netherlands Organization for Scientific Research (NWO). YW is also supported by the ERC Consolidator Grant STRINGFLATION under the HORIZON 2020 grant agreement no. 647995.
We acknowledge the hospitality of the Lorentz Center where this project was started.
\end{acknowledgements}

\appendix

\section{Algebraic Relations}\label{app:algebraicrelations}
There are a few identities we can use to rewrite our expressions for the slow-roll parameters, turn rate, etcetera. These are the scaling relation of the potential and the integration constant that quickly decays to zero
\begin{align}
 & X V_X -R_0 V_Y =2 \beta V, \label{eqn:relationpotential} \\
 & \dot X X e^{2Y/R_0}-R_0\dot Y +2\beta H  = 0\ , \label{eqn:integrationconstant}
\end{align}
together with the field equations of motion
\begin{align}
&\ddot X + 3 H \dot X + \frac{2}{R_0} \dot Y \dot X + e^{- 2 Y / R_0  }V_X = 0 , \\
&\ddot Y + 3 H \dot Y - \frac{1}{R_0} e^{ 2 Y / R_0  } \dot X^2 +V_Y = 0 .
\end{align}
We wish to find expressions that allow us to parameterize deviations from the ultra-light scenario, in which $Y$ is constant.
Therefore, for notational convenience, we define
\begin{align}
 \epsilon_Y& \equiv \frac{\dot Y^2}{2H^2}, \\
  \eta_Y &\equiv \frac{\dot \epsilon_Y}{H \epsilon_Y},\\
  \xi_Y &\equiv \frac{\dot \eta_Y}{H \eta_Y}, \\
  \left(1 +\Delta\right)^2 &= 1+ \frac{X^2}{R_0^2}e^{2Y/R_0} -\frac{2\beta^2}{\epsilon R_0^2}\ . \label{eqn:deltadef}
\end{align}
 Here $\Delta$ appears in the gauge transformation \eqref{eqn:gaugetransformationsigma}, and determines both the turn rate \eqref{eqn:turnrate} and the entropy mass of the isocurvature perturbations \eqref{eqn:entropymass}. Notice that $\epsilon_Y$ is simply the contribution of the kinetic energy of $Y$ to $\epsilon$, and therefore by definition $\epsilon_Y \leq \epsilon$. In the ultra-light scenario all these parameters are zero. We will see in a moment how to take the limit properly. \\

We can eliminate all $\dot X$ in terms of $\dot Y$ using \eqref{eqn:integrationconstant} and all $V_X$ in terms of $V_Y$ using \eqref{eqn:relationpotential}. This allows us to solve for the combination that appears in the definition \eqref{eqn:deltadef}
\begin{equation}
\frac{X^2 e^{2Y/R_0}}{2\beta^2} = \frac{\left(1-{\dot Y R_0}/({2\beta H})\right)^2}{\epsilon - \epsilon_Y} , \label{eqn:expressioneps0}
\end{equation}
which reduces to $1/\epsilon$ when $\dot Y=0$, such that $\Delta$ becomes zero. Indeed, we can use this result to rewrite
\begin{equation}
 (1+ \Delta)^2 =\frac{\left[1-{\beta \dot Y}/{(\epsilon R_0 H)}\right]^2}{{1-{\dot Y^2}/{(2\epsilon H^2)}}}\ .
\end{equation}
Having found this expression, we can now compute the physical quantities, such as the turn rate and the entropy mass along the inflationary trajectory. \\

First of all, the turn rate is given by
\begin{equation}
 \frac{\Omega^2}{H^2} = \frac{\beta^2 \eta^2 }{2\epsilon R^2_0}\frac{1-{\dot Y^2}/{(2\epsilon H^2)}}{\left[1-{\beta\dot Y}/{(\epsilon R_0 H)}\right]^2},
\end{equation}
which decreases with $\dot Y$, provided that the other parameters are kept constant. Moreover, taking a time derivative of \eqref{eqn:expressioneps0} we can relate $\eta$ to $\epsilon$.
In the ultra-light regime this reduces to the simple expression
\begin{equation}
 \eta=\frac{2\epsilon}{\beta} \quad \text{if} \quad \dot Y =0\ .
\end{equation}

It is a bit more involved to find the expression of the entropy mass. First we need to compute the time derivative of $\Delta$:
\begin{align}
 \frac{\dot \Delta}{1+\Delta}&  =  \frac{\beta \dot Y}{4\epsilon R_0 H}\frac{H}{1-{\beta \dot Y}/{(\epsilon R_0 H)}} \nonumber \\
 & ~~  \times \left[2\eta + (\eta-\eta_Y)
 \frac{2-{\dot Y R_0}/{(\beta H)}}{1-{\dot Y^2}/{(2\epsilon H^2)}}\right]
\end{align}
Notice that we have rewritten the time derivatives of $\epsilon_Y$ in terms of $\eta_Y$. We need to do this to have an expression that goes manifestly to zero in the ultra-light limit, which is otherwise hidden by the the non-trivial relation between $\eta$ and $\epsilon$ for the general background solution. From (\ref{eqn:entropymass}) that the entropy mass is given by
\begin{equation}
 \mu^2 = -3H\frac{\dot \Delta}{1+\Delta} -\left(\frac{\dot \Delta}{1+\Delta} \right)^2 - \partial_t \left(\frac{\dot \Delta}{1+\Delta} \right)\ ,
\end{equation}
which is a functional of
$\dot Y$, ${\epsilon_Y}/{\epsilon}$, $\eta_Y$, $\xi_Y$, $\eta$ and $\xi$.
Interestingly, the above expression has the same form with the entropy mass derived from the Hamilton-Jacobi formalism, after identifying $\frac{\dot \Delta}{1+\Delta} \to - W_{NN}$, where $W_{NN}$ is the Hessian of the fake superpotential $W$ projected along the direction orthogonal to the inflationary path~\cite{Achucarro:2018ngj}. The possible relation with the multi-field Hamilton-Jacobi system deserves a closer look. In addition, the entropy mass vanishes for a constant $\Delta$, which is the case for the isometry trajectories ($\Delta=0$) and $V_Y\simeq0$ attractors ($\Delta=-2\beta/(3R_0^2)$).  \\

\section{Gauge transformation}
\label{sec:gauge}

In this Appendix, we derive the non-linear gauge transformation between the comoving gauge and ultra-light gauge, and build the second order relation from  $\F$ and $\varphi$ to the comoving curvature perturbation $\R$ and isocurvature perturbation $\sigma$. \\

Let us denote the spacetime coordinates in the ultra-light gauge as $\tilde x$ and in the comoving gauge as $x$, and the same for the fields. A scalar field transforms as a scalar under a gauge transformation, i.e.:
\begin{equation} \label{scalar}
 \tilde\phi^a(\tilde x) = \phi^a(x).
\end{equation}
In the ultra-light gauge the field fluctuations are given by $\tilde\phi^a(\tilde x)= \left(  e^{-\F(\tilde x)/R_0}{\X}(\tilde x), ~ Y(\tilde  x)+\F(\tilde x)\right)$. Writing $\tilde t = t + T(t,x)$, we can expand $\tilde\phi^a(\tilde x)$ to second order
\begin{widetext}
\begin{align}
\tilde\phi^a(\tilde x) =  \bigg(&X +\dot X T + \tfrac{1}{2}\ddot X T^2 - X \frac{\F}{R_0} - \dot X T \frac{\F}{R_0} - X \frac{\dot \F}{R_0} T + X \frac{\F^2}{2R_0^2}  \ ,\nn  \\
&Y + \dot Y T + \tfrac{1}{2}\ddot Y T^2 +\F + \dot\F T  \ \bigg).
\label{phigonzalogauge}
\end{align}
\end{widetext}
The comoving gauge is defined to have no perturbations along the inflaton trajectory.
At linear order this simply means the perturbation along the tangent direction vanishes and the one in the normal direction is the isocurvature mode $\sigma$.
Here we define the tangent and normal unit vectors as $T^a=(\dot X, ~ \dot Y)/\dot\phi$ and $N^a = (-\dot Y e^{-Y/R_0}, ~ \dot X e^{Y/R_0})/\dot\phi$, where $\dot\phi^2=e^{2Y/R_0}\dot X^2 +\dot Y^2$. Then at the leading order the scalar fields and perturbations are expressed as
 $\phi^a(x)=\phi_0^a(t)+\sigma (x)N^a(t)$.
Beyond linear order, the effects of the curved field manifold becomes nontrivial, and one convenient way is to formulate the field fluctuations as the one along a geodesic in the field space.
In our case, $\phi_0^a(t)$ and $N^a$ specify one geodesic along the normal direction.
Thus we can follow the covariant approach developed in Ref. \cite{Gong:2011uw, Gong:2016qmq} to parametrize the isocurvature perturbation up to second order $\phi^a(x)=\phi_0^a(t) +\sigma (x)N^a(t) -\frac{1}{2} \Gamma^a_{bc}N^bN^c \sigma^2$, which gives us
\begin{widetext}
\begin{align}
\phi^a(x) =\bigg(X + \sigma N^X - N^X N^Y \frac{\sigma^2}{R_0}
~ , ~ Y  + \sigma N^Y  + \frac{1}{2} e^{2Y/R_0}N_X^2\frac{\sigma^2}{R_0}
\bigg).
\label{phicomovinggauge}
\end{align}
\end{widetext}

Next, solving Eq.~\eqref{scalar} to first order we find
\begin{equation}
 T^{(1)}  = -\frac{\beta}{\epsilon H R_0} \F~,
\end{equation}
and also the relation between isocurvature perturbation and $\F$
\be
\sigma = (1+\Delta)\F ~,
\ee
where we used the identity \eqref{eqn:integrationconstant} and the parameter $\Delta$ in \eqref{eqn:deltadef} to simplify the expression.
After some algebra, the second order solution gives us
\begin{align}
 T^{(2)}
 &  =  \left(-\frac{\eta H}{4}+\frac{\epsilon H}{2}+\frac{\dot X_0^2 \dot Y_0 G_{XX}}{\epsilon H^2 R_0}\right)\frac{\beta^2}{\epsilon^2}\frac{\F^2}{ H^2 R_0^2} \nn\\
 & \quad  -\frac{X_0 \dot X_0 G_{XX}}{4\epsilon R_0^2
 H^2}\F^2- \frac{G_{XX} \dot X_0^2}{2\epsilon R_0 H^2}\left(\frac{\beta}{\epsilon H R_0}\right)\F^2 \nn\\
& \quad +\frac{\dot Y_0}{8\epsilon^2 R_0 H^4}\left(2\dot\phi_0^2-\dot Y_0^2\right)\sigma^2 +\frac{\beta^2}{\epsilon^2}\frac{\F \dot \F}{ H^2 R_0^2}~.
\end{align}

Meanwhile, the metric perturbations in these two gauges are connected by $a(t)e^{\R(t)}  = a(\tilde t) e^{\varphi(\tilde t)}$, which leads to
\be
\R + \frac{1}{2}\R^2 = \varphi + H T + \frac{1}{2}\varphi^2 + \dot\varphi T   + \frac{1}{2}\frac{\ddot a}{a} T^2 + HT \varphi ...
\ee
Here we neglected the spatial gradient terms \cite{Maldacena:2002vr}, since they are suppressed on superhorizon scales.
To first order we find
\begin{equation}
 \R^{(1)} = \varphi + H T^{(1)} =  \varphi -\frac{\beta}{\epsilon  R_0} \F~.
\end{equation}
And finally at the second order, the relation between curvature perturbation and two ultra-light fields $\F$ and $\varphi$ is  given by
\begin{widetext}
\begin{align}
 \R^{(2)}
& =  H T^{(2)} -\frac{\beta}{\epsilon} \frac{\dot\varphi}{H} \frac{\F}{R_0}   - \frac{\beta^2}{2\epsilon} \frac{\F^2}{R_0^2} \nn\\
& =
-\left[\frac{\eta}{4} + \frac{\epsilon}{2 \beta} +\frac{\epsilon R_0}{4\beta^2}\frac{\dot Y}{H}-\frac{\dot Y^2}{2 \beta H^2}  -\frac{2\dot Y}{R_0 H}\left(1-\frac{\dot Y^2}{2\epsilon H^2}\right)
-\frac{\epsilon R_0}{4\beta^2}\frac{\dot Y}{H}\left(4-\frac{\dot Y^2}{\epsilon H^2}\right)(1+\Delta)^2\right]\frac{\beta^2}{\epsilon^2}\frac{\F^2}{R_0^2} \nn\\
& \quad -\frac{\beta^2}{\epsilon^2} \frac{\F \dot \F}{R_0^2 H} -\frac{\beta}{\epsilon} \frac{\dot\varphi}{H} \frac{\F}{R_0} ~.
\end{align}
\end{widetext}
On superhorizon scales, since $\F$ and $\varphi$ freeze, we can drop the last two terms with time derivatives.

\bibliography{bibfile}

\begin{thebibliography}{48}%
\makeatletter
\providecommand \@ifxundefined [1]{%
 \@ifx{#1\undefined}
}%
\providecommand \@ifnum [1]{%
 \ifnum #1\expandafter \@firstoftwo
 \else \expandafter \@secondoftwo
 \fi
}%
\providecommand \@ifx [1]{%
 \ifx #1\expandafter \@firstoftwo
 \else \expandafter \@secondoftwo
 \fi
}%
\providecommand \natexlab [1]{#1}%
\providecommand \enquote  [1]{``#1''}%
\providecommand \bibnamefont  [1]{#1}%
\providecommand \bibfnamefont [1]{#1}%
\providecommand \citenamefont [1]{#1}%
\providecommand \href@noop [0]{\@secondoftwo}%
\providecommand \href [0]{\begingroup \@sanitize@url \@href}%
\providecommand \@href[1]{\@@startlink{#1}\@@href}%
\providecommand \@@href[1]{\endgroup#1\@@endlink}%
\providecommand \@sanitize@url [0]{\catcode `\\12\catcode `\$12\catcode
  `\&12\catcode `\#12\catcode `\^12\catcode `\_12\catcode `\%12\relax}%
\providecommand \@@startlink[1]{}%
\providecommand \@@endlink[0]{}%
\providecommand \url  [0]{\begingroup\@sanitize@url \@url }%
\providecommand \@url [1]{\endgroup\@href {#1}{\urlprefix }}%
\providecommand \urlprefix  [0]{URL }%
\providecommand \Eprint [0]{\href }%
\providecommand \doibase [0]{http://dx.doi.org/}%
\providecommand \selectlanguage [0]{\@gobble}%
\providecommand \bibinfo  [0]{\@secondoftwo}%
\providecommand \bibfield  [0]{\@secondoftwo}%
\providecommand \translation [1]{[#1]}%
\providecommand \BibitemOpen [0]{}%
\providecommand \bibitemStop [0]{}%
\providecommand \bibitemNoStop [0]{.\EOS\space}%
\providecommand \EOS [0]{\spacefactor3000\relax}%
\providecommand \BibitemShut  [1]{\csname bibitem#1\endcsname}%
\let\auto@bib@innerbib\@empty
\bibitem [{\citenamefont {Ach\'{u}carro}\ \emph {et~al.}(2017)\citenamefont
  {Ach\'{u}carro}, \citenamefont {Atal}, \citenamefont {Germani},\ and\
  \citenamefont {Palma}}]{Achucarro:2016fby}%
  \BibitemOpen
  \bibfield  {author} {\bibinfo {author} {\bibfnamefont {A.}~\bibnamefont
  {Ach\'{u}carro}}, \bibinfo {author} {\bibfnamefont {V.}~\bibnamefont {Atal}},
  \bibinfo {author} {\bibfnamefont {C.}~\bibnamefont {Germani}}, \ and\
  \bibinfo {author} {\bibfnamefont {G.~A.}\ \bibnamefont {Palma}},\ }\href
  {\doibase 10.1088/1475-7516/2017/02/013} {\bibfield  {journal} {\bibinfo
  {journal} {JCAP}\ }\textbf {\bibinfo {volume} {1702}},\ \bibinfo {pages}
  {013} (\bibinfo {year} {2017})},\ \Eprint {http://arxiv.org/abs/1607.08609}
  {arXiv:1607.08609 [astro-ph.CO]} \BibitemShut {NoStop}%
\bibitem [{\citenamefont {Ach\'ucarro}\ \emph
  {et~al.}(2019{\natexlab{a}})\citenamefont {Ach\'ucarro}, \citenamefont
  {Copeland}, \citenamefont {Iarygina}, \citenamefont {Palma}, \citenamefont
  {Wang},\ and\ \citenamefont {Welling}}]{Achucarro:2019pux}%
  \BibitemOpen
  \bibfield  {author} {\bibinfo {author} {\bibfnamefont {A.}~\bibnamefont
  {Ach\'ucarro}}, \bibinfo {author} {\bibfnamefont {E.~J.}\ \bibnamefont
  {Copeland}}, \bibinfo {author} {\bibfnamefont {O.}~\bibnamefont {Iarygina}},
  \bibinfo {author} {\bibfnamefont {G.~A.}\ \bibnamefont {Palma}}, \bibinfo
  {author} {\bibfnamefont {D.-G.}\ \bibnamefont {Wang}}, \ and\ \bibinfo
  {author} {\bibfnamefont {Y.}~\bibnamefont {Welling}},\ }\href@noop {} {\
  (\bibinfo {year} {2019}{\natexlab{a}})},\ \Eprint
  {http://arxiv.org/abs/1901.03657} {arXiv:1901.03657 [astro-ph.CO]}
  \BibitemShut {NoStop}%
\bibitem [{\citenamefont {Enqvist}\ and\ \citenamefont
  {Vaihkonen}(2004)}]{Enqvist:2004bk}%
  \BibitemOpen
  \bibfield  {author} {\bibinfo {author} {\bibfnamefont {K.}~\bibnamefont
  {Enqvist}}\ and\ \bibinfo {author} {\bibfnamefont {A.}~\bibnamefont
  {Vaihkonen}},\ }\href {\doibase 10.1088/1475-7516/2004/09/006} {\bibfield
  {journal} {\bibinfo  {journal} {JCAP}\ }\textbf {\bibinfo {volume} {0409}},\
  \bibinfo {pages} {006} (\bibinfo {year} {2004})},\ \Eprint
  {http://arxiv.org/abs/hep-ph/0405103} {arXiv:hep-ph/0405103 [hep-ph]}
  \BibitemShut {NoStop}%
\bibitem [{\citenamefont {Lyth}\ and\ \citenamefont
  {Rodriguez}(2005)}]{Lyth:2005fi}%
  \BibitemOpen
  \bibfield  {author} {\bibinfo {author} {\bibfnamefont {D.~H.}\ \bibnamefont
  {Lyth}}\ and\ \bibinfo {author} {\bibfnamefont {Y.}~\bibnamefont
  {Rodriguez}},\ }\href {\doibase 10.1103/PhysRevLett.95.121302} {\bibfield
  {journal} {\bibinfo  {journal} {Phys. Rev. Lett.}\ }\textbf {\bibinfo
  {volume} {95}},\ \bibinfo {pages} {121302} (\bibinfo {year} {2005})},\
  \Eprint {http://arxiv.org/abs/astro-ph/0504045} {arXiv:astro-ph/0504045
  [astro-ph]} \BibitemShut {NoStop}%
\bibitem [{\citenamefont {Seery}\ and\ \citenamefont
  {Lidsey}(2005)}]{Seery:2005gb}%
  \BibitemOpen
  \bibfield  {author} {\bibinfo {author} {\bibfnamefont {D.}~\bibnamefont
  {Seery}}\ and\ \bibinfo {author} {\bibfnamefont {J.~E.}\ \bibnamefont
  {Lidsey}},\ }\href {\doibase 10.1088/1475-7516/2005/09/011} {\bibfield
  {journal} {\bibinfo  {journal} {JCAP}\ }\textbf {\bibinfo {volume} {0509}},\
  \bibinfo {pages} {011} (\bibinfo {year} {2005})},\ \Eprint
  {http://arxiv.org/abs/astro-ph/0506056} {arXiv:astro-ph/0506056 [astro-ph]}
  \BibitemShut {NoStop}%
\bibitem [{\citenamefont {Rigopoulos}\ \emph {et~al.}(2006)\citenamefont
  {Rigopoulos}, \citenamefont {Shellard},\ and\ \citenamefont {van
  Tent}}]{Rigopoulos:2005ae}%
  \BibitemOpen
  \bibfield  {author} {\bibinfo {author} {\bibfnamefont {G.~I.}\ \bibnamefont
  {Rigopoulos}}, \bibinfo {author} {\bibfnamefont {E.~P.~S.}\ \bibnamefont
  {Shellard}}, \ and\ \bibinfo {author} {\bibfnamefont {B.~J.~W.}\ \bibnamefont
  {van Tent}},\ }\href {\doibase 10.1103/PhysRevD.73.083522} {\bibfield
  {journal} {\bibinfo  {journal} {Phys. Rev.}\ }\textbf {\bibinfo {volume}
  {D73}},\ \bibinfo {pages} {083522} (\bibinfo {year} {2006})},\ \Eprint
  {http://arxiv.org/abs/astro-ph/0506704} {arXiv:astro-ph/0506704 [astro-ph]}
  \BibitemShut {NoStop}%
\bibitem [{\citenamefont {Alabidi}\ and\ \citenamefont
  {Lyth}(2006)}]{Alabidi:2005qi}%
  \BibitemOpen
  \bibfield  {author} {\bibinfo {author} {\bibfnamefont {L.}~\bibnamefont
  {Alabidi}}\ and\ \bibinfo {author} {\bibfnamefont {D.~H.}\ \bibnamefont
  {Lyth}},\ }\href {\doibase 10.1088/1475-7516/2006/05/016} {\bibfield
  {journal} {\bibinfo  {journal} {JCAP}\ }\textbf {\bibinfo {volume} {0605}},\
  \bibinfo {pages} {016} (\bibinfo {year} {2006})},\ \Eprint
  {http://arxiv.org/abs/astro-ph/0510441} {arXiv:astro-ph/0510441 [astro-ph]}
  \BibitemShut {NoStop}%
\bibitem [{\citenamefont {Battefeld}\ and\ \citenamefont
  {Battefeld}(2007)}]{Battefeld:2007en}%
  \BibitemOpen
  \bibfield  {author} {\bibinfo {author} {\bibfnamefont {D.}~\bibnamefont
  {Battefeld}}\ and\ \bibinfo {author} {\bibfnamefont {T.}~\bibnamefont
  {Battefeld}},\ }\href {\doibase 10.1088/1475-7516/2007/05/012} {\bibfield
  {journal} {\bibinfo  {journal} {JCAP}\ }\textbf {\bibinfo {volume} {0705}},\
  \bibinfo {pages} {012} (\bibinfo {year} {2007})},\ \Eprint
  {http://arxiv.org/abs/hep-th/0703012} {arXiv:hep-th/0703012 [hep-th]}
  \BibitemShut {NoStop}%
\bibitem [{\citenamefont {Choi}\ \emph {et~al.}(2007)\citenamefont {Choi},
  \citenamefont {Hall},\ and\ \citenamefont {van~de Bruck}}]{Choi:2007su}%
  \BibitemOpen
  \bibfield  {author} {\bibinfo {author} {\bibfnamefont {K.-Y.}\ \bibnamefont
  {Choi}}, \bibinfo {author} {\bibfnamefont {L.~M.~H.}\ \bibnamefont {Hall}}, \
  and\ \bibinfo {author} {\bibfnamefont {C.}~\bibnamefont {van~de Bruck}},\
  }\href {\doibase 10.1088/1475-7516/2007/02/029} {\bibfield  {journal}
  {\bibinfo  {journal} {JCAP}\ }\textbf {\bibinfo {volume} {0702}},\ \bibinfo
  {pages} {029} (\bibinfo {year} {2007})},\ \Eprint
  {http://arxiv.org/abs/astro-ph/0701247} {arXiv:astro-ph/0701247 [astro-ph]}
  \BibitemShut {NoStop}%
\bibitem [{\citenamefont {Byrnes}\ \emph {et~al.}(2008)\citenamefont {Byrnes},
  \citenamefont {Choi},\ and\ \citenamefont {Hall}}]{Byrnes:2008wi}%
  \BibitemOpen
  \bibfield  {author} {\bibinfo {author} {\bibfnamefont {C.~T.}\ \bibnamefont
  {Byrnes}}, \bibinfo {author} {\bibfnamefont {K.-Y.}\ \bibnamefont {Choi}}, \
  and\ \bibinfo {author} {\bibfnamefont {L.~M.~H.}\ \bibnamefont {Hall}},\
  }\href {\doibase 10.1088/1475-7516/2008/10/008} {\bibfield  {journal}
  {\bibinfo  {journal} {JCAP}\ }\textbf {\bibinfo {volume} {0810}},\ \bibinfo
  {pages} {008} (\bibinfo {year} {2008})},\ \Eprint
  {http://arxiv.org/abs/0807.1101} {arXiv:0807.1101 [astro-ph]} \BibitemShut
  {NoStop}%
\bibitem [{\citenamefont {Byrnes}\ and\ \citenamefont
  {Tasinato}(2009)}]{Byrnes:2009qy}%
  \BibitemOpen
  \bibfield  {author} {\bibinfo {author} {\bibfnamefont {C.~T.}\ \bibnamefont
  {Byrnes}}\ and\ \bibinfo {author} {\bibfnamefont {G.}~\bibnamefont
  {Tasinato}},\ }\href {\doibase 10.1088/1475-7516/2009/08/016} {\bibfield
  {journal} {\bibinfo  {journal} {JCAP}\ }\textbf {\bibinfo {volume} {0908}},\
  \bibinfo {pages} {016} (\bibinfo {year} {2009})},\ \Eprint
  {http://arxiv.org/abs/0906.0767} {arXiv:0906.0767 [astro-ph.CO]} \BibitemShut
  {NoStop}%
\bibitem [{\citenamefont {Battefeld}\ and\ \citenamefont
  {Battefeld}(2009)}]{Battefeld:2009ym}%
  \BibitemOpen
  \bibfield  {author} {\bibinfo {author} {\bibfnamefont {D.}~\bibnamefont
  {Battefeld}}\ and\ \bibinfo {author} {\bibfnamefont {T.}~\bibnamefont
  {Battefeld}},\ }\href {\doibase 10.1088/1475-7516/2009/11/010} {\bibfield
  {journal} {\bibinfo  {journal} {JCAP}\ }\textbf {\bibinfo {volume} {0911}},\
  \bibinfo {pages} {010} (\bibinfo {year} {2009})},\ \Eprint
  {http://arxiv.org/abs/0908.4269} {arXiv:0908.4269 [hep-th]} \BibitemShut
  {NoStop}%
\bibitem [{\citenamefont {Chen}\ and\ \citenamefont
  {Wang}(2010{\natexlab{a}})}]{Chen:2009we}%
  \BibitemOpen
  \bibfield  {author} {\bibinfo {author} {\bibfnamefont {X.}~\bibnamefont
  {Chen}}\ and\ \bibinfo {author} {\bibfnamefont {Y.}~\bibnamefont {Wang}},\
  }\href {\doibase 10.1103/PhysRevD.81.063511} {\bibfield  {journal} {\bibinfo
  {journal} {Phys. Rev.}\ }\textbf {\bibinfo {volume} {D81}},\ \bibinfo {pages}
  {063511} (\bibinfo {year} {2010}{\natexlab{a}})},\ \Eprint
  {http://arxiv.org/abs/0909.0496} {arXiv:0909.0496 [astro-ph.CO]} \BibitemShut
  {NoStop}%
\bibitem [{\citenamefont {Chen}\ and\ \citenamefont
  {Wang}(2010{\natexlab{b}})}]{Chen:2009zp}%
  \BibitemOpen
  \bibfield  {author} {\bibinfo {author} {\bibfnamefont {X.}~\bibnamefont
  {Chen}}\ and\ \bibinfo {author} {\bibfnamefont {Y.}~\bibnamefont {Wang}},\
  }\href {\doibase 10.1088/1475-7516/2010/04/027} {\bibfield  {journal}
  {\bibinfo  {journal} {JCAP}\ }\textbf {\bibinfo {volume} {1004}},\ \bibinfo
  {pages} {027} (\bibinfo {year} {2010}{\natexlab{b}})},\ \Eprint
  {http://arxiv.org/abs/0911.3380} {arXiv:0911.3380 [hep-th]} \BibitemShut
  {NoStop}%
\bibitem [{\citenamefont {Elliston}\ \emph {et~al.}(2011)\citenamefont
  {Elliston}, \citenamefont {Mulryne}, \citenamefont {Seery},\ and\
  \citenamefont {Tavakol}}]{Elliston:2011et}%
  \BibitemOpen
  \bibfield  {author} {\bibinfo {author} {\bibfnamefont {J.}~\bibnamefont
  {Elliston}}, \bibinfo {author} {\bibfnamefont {D.}~\bibnamefont {Mulryne}},
  \bibinfo {author} {\bibfnamefont {D.}~\bibnamefont {Seery}}, \ and\ \bibinfo
  {author} {\bibfnamefont {R.}~\bibnamefont {Tavakol}},\ }\href {\doibase
  10.1142/S0217751X11054280, 10.1142/S2010194511001292} {\bibfield  {journal}
  {\bibinfo  {journal} {Int. J. Mod. Phys.}\ }\textbf {\bibinfo {volume}
  {A26}},\ \bibinfo {pages} {3821} (\bibinfo {year} {2011})},\ \bibinfo {note}
  {[Int. J. Mod. Phys. Conf. Ser.03,203(2011)]},\ \Eprint
  {http://arxiv.org/abs/1107.2270} {arXiv:1107.2270 [astro-ph.CO]} \BibitemShut
  {NoStop}%
\bibitem [{\citenamefont {Mulryne}\ \emph {et~al.}(2011)\citenamefont
  {Mulryne}, \citenamefont {Orani},\ and\ \citenamefont
  {Rajantie}}]{Mulryne:2011ni}%
  \BibitemOpen
  \bibfield  {author} {\bibinfo {author} {\bibfnamefont {D.}~\bibnamefont
  {Mulryne}}, \bibinfo {author} {\bibfnamefont {S.}~\bibnamefont {Orani}}, \
  and\ \bibinfo {author} {\bibfnamefont {A.}~\bibnamefont {Rajantie}},\ }\href
  {\doibase 10.1103/PhysRevD.84.123527} {\bibfield  {journal} {\bibinfo
  {journal} {Phys. Rev.}\ }\textbf {\bibinfo {volume} {D84}},\ \bibinfo {pages}
  {123527} (\bibinfo {year} {2011})},\ \Eprint {http://arxiv.org/abs/1107.4739}
  {arXiv:1107.4739 [hep-th]} \BibitemShut {NoStop}%
\bibitem [{\citenamefont {McAllister}\ \emph {et~al.}(2012)\citenamefont
  {McAllister}, \citenamefont {Renaux-Petel},\ and\ \citenamefont
  {Xu}}]{McAllister:2012am}%
  \BibitemOpen
  \bibfield  {author} {\bibinfo {author} {\bibfnamefont {L.}~\bibnamefont
  {McAllister}}, \bibinfo {author} {\bibfnamefont {S.}~\bibnamefont
  {Renaux-Petel}}, \ and\ \bibinfo {author} {\bibfnamefont {G.}~\bibnamefont
  {Xu}},\ }\href {\doibase 10.1088/1475-7516/2012/10/046} {\bibfield  {journal}
  {\bibinfo  {journal} {JCAP}\ }\textbf {\bibinfo {volume} {1210}},\ \bibinfo
  {pages} {046} (\bibinfo {year} {2012})},\ \Eprint
  {http://arxiv.org/abs/1207.0317} {arXiv:1207.0317 [astro-ph.CO]} \BibitemShut
  {NoStop}%
\bibitem [{\citenamefont {Byrnes}\ and\ \citenamefont
  {Gong}(2013)}]{Byrnes:2012sc}%
  \BibitemOpen
  \bibfield  {author} {\bibinfo {author} {\bibfnamefont {C.~T.}\ \bibnamefont
  {Byrnes}}\ and\ \bibinfo {author} {\bibfnamefont {J.-O.}\ \bibnamefont
  {Gong}},\ }\href {\doibase 10.1016/j.physletb.2012.11.052} {\bibfield
  {journal} {\bibinfo  {journal} {Phys. Lett.}\ }\textbf {\bibinfo {volume}
  {B718}},\ \bibinfo {pages} {718} (\bibinfo {year} {2013})},\ \Eprint
  {http://arxiv.org/abs/1210.1851} {arXiv:1210.1851 [astro-ph.CO]} \BibitemShut
  {NoStop}%
\bibitem [{\citenamefont {Baumann}\ and\ \citenamefont
  {Green}(2012)}]{Baumann:2011nk}%
  \BibitemOpen
  \bibfield  {author} {\bibinfo {author} {\bibfnamefont {D.}~\bibnamefont
  {Baumann}}\ and\ \bibinfo {author} {\bibfnamefont {D.}~\bibnamefont
  {Green}},\ }\href {\doibase 10.1103/PhysRevD.85.103520} {\bibfield  {journal}
  {\bibinfo  {journal} {Phys. Rev.}\ }\textbf {\bibinfo {volume} {D85}},\
  \bibinfo {pages} {103520} (\bibinfo {year} {2012})},\ \Eprint
  {http://arxiv.org/abs/1109.0292} {arXiv:1109.0292 [hep-th]} \BibitemShut
  {NoStop}%
\bibitem [{\citenamefont {Arkani-Hamed}\ and\ \citenamefont
  {Maldacena}(2015)}]{Arkani-Hamed:2015bza}%
  \BibitemOpen
  \bibfield  {author} {\bibinfo {author} {\bibfnamefont {N.}~\bibnamefont
  {Arkani-Hamed}}\ and\ \bibinfo {author} {\bibfnamefont {J.}~\bibnamefont
  {Maldacena}},\ }\href@noop {} {\  (\bibinfo {year} {2015})},\ \Eprint
  {http://arxiv.org/abs/1503.08043} {arXiv:1503.08043 [hep-th]} \BibitemShut
  {NoStop}%
\bibitem [{\citenamefont {Lee}\ \emph {et~al.}(2016)\citenamefont {Lee},
  \citenamefont {Baumann},\ and\ \citenamefont {Pimentel}}]{Lee:2016vti}%
  \BibitemOpen
  \bibfield  {author} {\bibinfo {author} {\bibfnamefont {H.}~\bibnamefont
  {Lee}}, \bibinfo {author} {\bibfnamefont {D.}~\bibnamefont {Baumann}}, \ and\
  \bibinfo {author} {\bibfnamefont {G.~L.}\ \bibnamefont {Pimentel}},\ }\href
  {\doibase 10.1007/JHEP12(2016)040} {\bibfield  {journal} {\bibinfo  {journal}
  {JHEP}\ }\textbf {\bibinfo {volume} {12}},\ \bibinfo {pages} {040} (\bibinfo
  {year} {2016})},\ \Eprint {http://arxiv.org/abs/1607.03735} {arXiv:1607.03735
  [hep-th]} \BibitemShut {NoStop}%
\bibitem [{\citenamefont {Chen}\ \emph
  {et~al.}(2018{\natexlab{a}})\citenamefont {Chen}, \citenamefont {Palma},
  \citenamefont {Riquelme}, \citenamefont {Scheihing~Hitschfeld},\ and\
  \citenamefont {Sypsas}}]{Chen:2018uul}%
  \BibitemOpen
  \bibfield  {author} {\bibinfo {author} {\bibfnamefont {X.}~\bibnamefont
  {Chen}}, \bibinfo {author} {\bibfnamefont {G.~A.}\ \bibnamefont {Palma}},
  \bibinfo {author} {\bibfnamefont {W.}~\bibnamefont {Riquelme}}, \bibinfo
  {author} {\bibfnamefont {B.}~\bibnamefont {Scheihing~Hitschfeld}}, \ and\
  \bibinfo {author} {\bibfnamefont {S.}~\bibnamefont {Sypsas}},\ }\href
  {\doibase 10.1103/PhysRevD.98.083528} {\bibfield  {journal} {\bibinfo
  {journal} {Phys. Rev.}\ }\textbf {\bibinfo {volume} {D98}},\ \bibinfo {pages}
  {083528} (\bibinfo {year} {2018}{\natexlab{a}})},\ \Eprint
  {http://arxiv.org/abs/1804.07315} {arXiv:1804.07315 [hep-th]} \BibitemShut
  {NoStop}%
\bibitem [{\citenamefont {Chen}\ \emph
  {et~al.}(2018{\natexlab{b}})\citenamefont {Chen}, \citenamefont {Palma},
  \citenamefont {Scheihing~Hitschfeld},\ and\ \citenamefont
  {Sypsas}}]{Chen:2018brw}%
  \BibitemOpen
  \bibfield  {author} {\bibinfo {author} {\bibfnamefont {X.}~\bibnamefont
  {Chen}}, \bibinfo {author} {\bibfnamefont {G.~A.}\ \bibnamefont {Palma}},
  \bibinfo {author} {\bibfnamefont {B.}~\bibnamefont {Scheihing~Hitschfeld}}, \
  and\ \bibinfo {author} {\bibfnamefont {S.}~\bibnamefont {Sypsas}},\ }\href
  {\doibase 10.1103/PhysRevLett.121.161302} {\bibfield  {journal} {\bibinfo
  {journal} {Phys. Rev. Lett.}\ }\textbf {\bibinfo {volume} {121}},\ \bibinfo
  {pages} {161302} (\bibinfo {year} {2018}{\natexlab{b}})},\ \Eprint
  {http://arxiv.org/abs/1806.05202} {arXiv:1806.05202 [astro-ph.CO]}
  \BibitemShut {NoStop}%
\bibitem [{\citenamefont {Garcia-Saenz}\ \emph {et~al.}(2018)\citenamefont
  {Garcia-Saenz}, \citenamefont {Renaux-Petel},\ and\ \citenamefont
  {Ronayne}}]{Garcia-Saenz:2018ifx}%
  \BibitemOpen
  \bibfield  {author} {\bibinfo {author} {\bibfnamefont {S.}~\bibnamefont
  {Garcia-Saenz}}, \bibinfo {author} {\bibfnamefont {S.}~\bibnamefont
  {Renaux-Petel}}, \ and\ \bibinfo {author} {\bibfnamefont {J.}~\bibnamefont
  {Ronayne}},\ }\href {\doibase 10.1088/1475-7516/2018/07/057} {\bibfield
  {journal} {\bibinfo  {journal} {JCAP}\ }\textbf {\bibinfo {volume} {1807}},\
  \bibinfo {pages} {057} (\bibinfo {year} {2018})},\ \Eprint
  {http://arxiv.org/abs/1804.11279} {arXiv:1804.11279 [astro-ph.CO]}
  \BibitemShut {NoStop}%
\bibitem [{\citenamefont {Fumagalli}\ \emph {et~al.}(2019)\citenamefont
  {Fumagalli}, \citenamefont {Garcia-Saenz}, \citenamefont {Pinol},
  \citenamefont {Renaux-Petel},\ and\ \citenamefont
  {Ronayne}}]{Fumagalli:2019noh}%
  \BibitemOpen
  \bibfield  {author} {\bibinfo {author} {\bibfnamefont {J.}~\bibnamefont
  {Fumagalli}}, \bibinfo {author} {\bibfnamefont {S.}~\bibnamefont
  {Garcia-Saenz}}, \bibinfo {author} {\bibfnamefont {L.}~\bibnamefont {Pinol}},
  \bibinfo {author} {\bibfnamefont {S.}~\bibnamefont {Renaux-Petel}}, \ and\
  \bibinfo {author} {\bibfnamefont {J.}~\bibnamefont {Ronayne}},\ }\href@noop
  {} {\  (\bibinfo {year} {2019})},\ \Eprint {http://arxiv.org/abs/1902.03221}
  {arXiv:1902.03221 [hep-th]} \BibitemShut {NoStop}%
\bibitem [{\citenamefont {Achúcarro}\ and\ \citenamefont
  {Welling}(2019)}]{Achucarro:2019mea}%
  \BibitemOpen
  \bibfield  {author} {\bibinfo {author} {\bibfnamefont {A.}~\bibnamefont
  {Achúcarro}}\ and\ \bibinfo {author} {\bibfnamefont {Y.}~\bibnamefont
  {Welling}},\ }\href@noop {} {\  (\bibinfo {year} {2019})},\ \Eprint
  {http://arxiv.org/abs/1907.02020} {arXiv:1907.02020 [hep-th]} \BibitemShut
  {NoStop}%
\bibitem [{\citenamefont {Welling}(2019)}]{Welling:2019bib}%
  \BibitemOpen
  \bibfield  {author} {\bibinfo {author} {\bibfnamefont {Y.}~\bibnamefont
  {Welling}},\ }\href@noop {} {\  (\bibinfo {year} {2019})},\ \Eprint
  {http://arxiv.org/abs/1907.02951} {arXiv:1907.02951 [astro-ph.CO]}
  \BibitemShut {NoStop}%
\bibitem [{\citenamefont {Garcia-Saenz}\ \emph {et~al.}(2019)\citenamefont
  {Garcia-Saenz}, \citenamefont {Pinol},\ and\ \citenamefont
  {Renaux-Petel}}]{Garcia-Saenz:2019njm}%
  \BibitemOpen
  \bibfield  {author} {\bibinfo {author} {\bibfnamefont {S.}~\bibnamefont
  {Garcia-Saenz}}, \bibinfo {author} {\bibfnamefont {L.}~\bibnamefont {Pinol}},
  \ and\ \bibinfo {author} {\bibfnamefont {S.}~\bibnamefont {Renaux-Petel}},\
  }\href@noop {} {\  (\bibinfo {year} {2019})},\ \Eprint
  {http://arxiv.org/abs/1907.10403} {arXiv:1907.10403 [hep-th]} \BibitemShut
  {NoStop}%
\bibitem [{\citenamefont {Guth}(1981)}]{Guth:1980zm}%
  \BibitemOpen
  \bibfield  {author} {\bibinfo {author} {\bibfnamefont {A.~H.}\ \bibnamefont
  {Guth}},\ }\href {\doibase 10.1103/PhysRevD.23.347} {\bibfield  {journal}
  {\bibinfo  {journal} {Phys. Rev.}\ }\textbf {\bibinfo {volume} {D23}},\
  \bibinfo {pages} {347} (\bibinfo {year} {1981})}\BibitemShut {NoStop}%
\bibitem [{\citenamefont {Starobinsky}(1980)}]{Starobinsky:1980te}%
  \BibitemOpen
  \bibfield  {author} {\bibinfo {author} {\bibfnamefont {A.~A.}\ \bibnamefont
  {Starobinsky}},\ }\href {\doibase 10.1016/0370-2693(80)90670-X} {\bibfield
  {journal} {\bibinfo  {journal} {Phys. Lett.}\ }\textbf {\bibinfo {volume}
  {B91}},\ \bibinfo {pages} {99} (\bibinfo {year} {1980})},\ \bibinfo {note}
  {[,771(1980)]}\BibitemShut {NoStop}%
\bibitem [{\citenamefont {Linde}(1982)}]{Linde:1981mu}%
  \BibitemOpen
  \bibfield  {author} {\bibinfo {author} {\bibfnamefont {A.~D.}\ \bibnamefont
  {Linde}},\ }\href {\doibase 10.1016/0370-2693(82)91219-9} {\bibfield
  {journal} {\bibinfo  {journal} {Phys. Lett.}\ }\textbf {\bibinfo {volume}
  {108B}},\ \bibinfo {pages} {389} (\bibinfo {year} {1982})}\BibitemShut
  {NoStop}%
\bibitem [{\citenamefont {Albrecht}\ and\ \citenamefont
  {Steinhardt}(1982)}]{Albrecht:1982wi}%
  \BibitemOpen
  \bibfield  {author} {\bibinfo {author} {\bibfnamefont {A.}~\bibnamefont
  {Albrecht}}\ and\ \bibinfo {author} {\bibfnamefont {P.~J.}\ \bibnamefont
  {Steinhardt}},\ }\href {\doibase 10.1103/PhysRevLett.48.1220} {\bibfield
  {journal} {\bibinfo  {journal} {Phys. Rev. Lett.}\ }\textbf {\bibinfo
  {volume} {48}},\ \bibinfo {pages} {1220} (\bibinfo {year}
  {1982})}\BibitemShut {NoStop}%
\bibitem [{\citenamefont {Mukhanov}\ and\ \citenamefont
  {Chibisov}(1981)}]{Mukhanov:1981xt}%
  \BibitemOpen
  \bibfield  {author} {\bibinfo {author} {\bibfnamefont {V.~F.}\ \bibnamefont
  {Mukhanov}}\ and\ \bibinfo {author} {\bibfnamefont {G.~V.}\ \bibnamefont
  {Chibisov}},\ }\href@noop {} {\bibfield  {journal} {\bibinfo  {journal} {JETP
  Lett.}\ }\textbf {\bibinfo {volume} {33}},\ \bibinfo {pages} {532} (\bibinfo
  {year} {1981})}\BibitemShut {NoStop}%
\bibitem [{\citenamefont {Gordon}\ \emph {et~al.}(2001)\citenamefont {Gordon},
  \citenamefont {Wands}, \citenamefont {Bassett},\ and\ \citenamefont
  {Maartens}}]{Gordon:2000hv}%
  \BibitemOpen
  \bibfield  {author} {\bibinfo {author} {\bibfnamefont {C.}~\bibnamefont
  {Gordon}}, \bibinfo {author} {\bibfnamefont {D.}~\bibnamefont {Wands}},
  \bibinfo {author} {\bibfnamefont {B.~A.}\ \bibnamefont {Bassett}}, \ and\
  \bibinfo {author} {\bibfnamefont {R.}~\bibnamefont {Maartens}},\ }\href
  {\doibase 10.1103/PhysRevD.63.023506} {\bibfield  {journal} {\bibinfo
  {journal} {Phys. Rev.}\ }\textbf {\bibinfo {volume} {D63}},\ \bibinfo {pages}
  {023506} (\bibinfo {year} {2001})},\ \Eprint
  {http://arxiv.org/abs/astro-ph/0009131} {arXiv:astro-ph/0009131 [astro-ph]}
  \BibitemShut {NoStop}%
\bibitem [{\citenamefont {Groot~Nibbelink}\ and\ \citenamefont {van
  Tent}(2000)}]{GrootNibbelink:2000vx}%
  \BibitemOpen
  \bibfield  {author} {\bibinfo {author} {\bibfnamefont {S.}~\bibnamefont
  {Groot~Nibbelink}}\ and\ \bibinfo {author} {\bibfnamefont {B.~J.~W.}\
  \bibnamefont {van Tent}},\ }\href@noop {} {\  (\bibinfo {year} {2000})},\
  \Eprint {http://arxiv.org/abs/hep-ph/0011325} {arXiv:hep-ph/0011325 [hep-ph]}
  \BibitemShut {NoStop}%
\bibitem [{\citenamefont {Groot~Nibbelink}\ and\ \citenamefont {van
  Tent}(2002)}]{GrootNibbelink:2001qt}%
  \BibitemOpen
  \bibfield  {author} {\bibinfo {author} {\bibfnamefont {S.}~\bibnamefont
  {Groot~Nibbelink}}\ and\ \bibinfo {author} {\bibfnamefont {B.~J.~W.}\
  \bibnamefont {van Tent}},\ }\href {\doibase 10.1088/0264-9381/19/4/302}
  {\bibfield  {journal} {\bibinfo  {journal} {Class. Quant. Grav.}\ }\textbf
  {\bibinfo {volume} {19}},\ \bibinfo {pages} {613} (\bibinfo {year} {2002})},\
  \Eprint {http://arxiv.org/abs/hep-ph/0107272} {arXiv:hep-ph/0107272 [hep-ph]}
  \BibitemShut {NoStop}%
\bibitem [{\citenamefont {Achucarro}\ \emph
  {et~al.}(2011{\natexlab{a}})\citenamefont {Achucarro}, \citenamefont {Gong},
  \citenamefont {Hardeman}, \citenamefont {Palma},\ and\ \citenamefont
  {Patil}}]{Achucarro:2010jv}%
  \BibitemOpen
  \bibfield  {author} {\bibinfo {author} {\bibfnamefont {A.}~\bibnamefont
  {Achucarro}}, \bibinfo {author} {\bibfnamefont {J.-O.}\ \bibnamefont {Gong}},
  \bibinfo {author} {\bibfnamefont {S.}~\bibnamefont {Hardeman}}, \bibinfo
  {author} {\bibfnamefont {G.~A.}\ \bibnamefont {Palma}}, \ and\ \bibinfo
  {author} {\bibfnamefont {S.~P.}\ \bibnamefont {Patil}},\ }\href {\doibase
  10.1103/PhysRevD.84.043502} {\bibfield  {journal} {\bibinfo  {journal} {Phys.
  Rev.}\ }\textbf {\bibinfo {volume} {D84}},\ \bibinfo {pages} {043502}
  (\bibinfo {year} {2011}{\natexlab{a}})},\ \Eprint
  {http://arxiv.org/abs/1005.3848} {arXiv:1005.3848 [hep-th]} \BibitemShut
  {NoStop}%
\bibitem [{\citenamefont {Achucarro}\ \emph
  {et~al.}(2011{\natexlab{b}})\citenamefont {Achucarro}, \citenamefont {Gong},
  \citenamefont {Hardeman}, \citenamefont {Palma},\ and\ \citenamefont
  {Patil}}]{Achucarro:2010da}%
  \BibitemOpen
  \bibfield  {author} {\bibinfo {author} {\bibfnamefont {A.}~\bibnamefont
  {Achucarro}}, \bibinfo {author} {\bibfnamefont {J.-O.}\ \bibnamefont {Gong}},
  \bibinfo {author} {\bibfnamefont {S.}~\bibnamefont {Hardeman}}, \bibinfo
  {author} {\bibfnamefont {G.~A.}\ \bibnamefont {Palma}}, \ and\ \bibinfo
  {author} {\bibfnamefont {S.~P.}\ \bibnamefont {Patil}},\ }\href {\doibase
  10.1088/1475-7516/2011/01/030} {\bibfield  {journal} {\bibinfo  {journal}
  {JCAP}\ }\textbf {\bibinfo {volume} {1101}},\ \bibinfo {pages} {030}
  (\bibinfo {year} {2011}{\natexlab{b}})},\ \Eprint
  {http://arxiv.org/abs/1010.3693} {arXiv:1010.3693 [hep-ph]} \BibitemShut
  {NoStop}%
\bibitem [{\citenamefont {Nicolis}\ and\ \citenamefont
  {Piazza}(2012)}]{Nicolis:2011pv}%
  \BibitemOpen
  \bibfield  {author} {\bibinfo {author} {\bibfnamefont {A.}~\bibnamefont
  {Nicolis}}\ and\ \bibinfo {author} {\bibfnamefont {F.}~\bibnamefont
  {Piazza}},\ }\href {\doibase 10.1007/JHEP06(2012)025} {\bibfield  {journal}
  {\bibinfo  {journal} {JHEP}\ }\textbf {\bibinfo {volume} {06}},\ \bibinfo
  {pages} {025} (\bibinfo {year} {2012})},\ \Eprint
  {http://arxiv.org/abs/1112.5174} {arXiv:1112.5174 [hep-th]} \BibitemShut
  {NoStop}%
\bibitem [{\citenamefont {Maldacena}(2003)}]{Maldacena:2002vr}%
  \BibitemOpen
  \bibfield  {author} {\bibinfo {author} {\bibfnamefont {J.~M.}\ \bibnamefont
  {Maldacena}},\ }\href {\doibase 10.1088/1126-6708/2003/05/013} {\bibfield
  {journal} {\bibinfo  {journal} {JHEP}\ }\textbf {\bibinfo {volume} {05}},\
  \bibinfo {pages} {013} (\bibinfo {year} {2003})},\ \Eprint
  {http://arxiv.org/abs/astro-ph/0210603} {arXiv:astro-ph/0210603 [astro-ph]}
  \BibitemShut {NoStop}%
\bibitem [{\citenamefont {Creminelli}\ and\ \citenamefont
  {Zaldarriaga}(2004)}]{Creminelli:2004yq}%
  \BibitemOpen
  \bibfield  {author} {\bibinfo {author} {\bibfnamefont {P.}~\bibnamefont
  {Creminelli}}\ and\ \bibinfo {author} {\bibfnamefont {M.}~\bibnamefont
  {Zaldarriaga}},\ }\href {\doibase 10.1088/1475-7516/2004/10/006} {\bibfield
  {journal} {\bibinfo  {journal} {JCAP}\ }\textbf {\bibinfo {volume} {0410}},\
  \bibinfo {pages} {006} (\bibinfo {year} {2004})},\ \Eprint
  {http://arxiv.org/abs/astro-ph/0407059} {arXiv:astro-ph/0407059 [astro-ph]}
  \BibitemShut {NoStop}%
\bibitem [{\citenamefont {Cheung}\ \emph {et~al.}(2008)\citenamefont {Cheung},
  \citenamefont {Fitzpatrick}, \citenamefont {Kaplan},\ and\ \citenamefont
  {Senatore}}]{Cheung:2007sv}%
  \BibitemOpen
  \bibfield  {author} {\bibinfo {author} {\bibfnamefont {C.}~\bibnamefont
  {Cheung}}, \bibinfo {author} {\bibfnamefont {A.~L.}\ \bibnamefont
  {Fitzpatrick}}, \bibinfo {author} {\bibfnamefont {J.}~\bibnamefont {Kaplan}},
  \ and\ \bibinfo {author} {\bibfnamefont {L.}~\bibnamefont {Senatore}},\
  }\href {\doibase 10.1088/1475-7516/2008/02/021} {\bibfield  {journal}
  {\bibinfo  {journal} {JCAP}\ }\textbf {\bibinfo {volume} {0802}},\ \bibinfo
  {pages} {021} (\bibinfo {year} {2008})},\ \Eprint
  {http://arxiv.org/abs/0709.0295} {arXiv:0709.0295 [hep-th]} \BibitemShut
  {NoStop}%
\bibitem [{\citenamefont {Bravo}\ \emph {et~al.}(2018)\citenamefont {Bravo},
  \citenamefont {Mooij}, \citenamefont {Palma},\ and\ \citenamefont
  {Pradenas}}]{Bravo:2017wyw}%
  \BibitemOpen
  \bibfield  {author} {\bibinfo {author} {\bibfnamefont {R.}~\bibnamefont
  {Bravo}}, \bibinfo {author} {\bibfnamefont {S.}~\bibnamefont {Mooij}},
  \bibinfo {author} {\bibfnamefont {G.~A.}\ \bibnamefont {Palma}}, \ and\
  \bibinfo {author} {\bibfnamefont {B.}~\bibnamefont {Pradenas}},\ }\href
  {\doibase 10.1088/1475-7516/2018/05/024} {\bibfield  {journal} {\bibinfo
  {journal} {JCAP}\ }\textbf {\bibinfo {volume} {1805}},\ \bibinfo {pages}
  {024} (\bibinfo {year} {2018})},\ \Eprint {http://arxiv.org/abs/1711.02680}
  {arXiv:1711.02680 [astro-ph.CO]} \BibitemShut {NoStop}%
\bibitem [{\citenamefont {Cabass}\ \emph {et~al.}(2018)\citenamefont {Cabass},
  \citenamefont {Pajer},\ and\ \citenamefont {van~der Woude}}]{Cabass:2018jgj}%
  \BibitemOpen
  \bibfield  {author} {\bibinfo {author} {\bibfnamefont {G.}~\bibnamefont
  {Cabass}}, \bibinfo {author} {\bibfnamefont {E.}~\bibnamefont {Pajer}}, \
  and\ \bibinfo {author} {\bibfnamefont {D.}~\bibnamefont {van~der Woude}},\
  }\href@noop {} {\  (\bibinfo {year} {2018})},\ \Eprint
  {http://arxiv.org/abs/1805.08775} {arXiv:1805.08775 [astro-ph.CO]}
  \BibitemShut {NoStop}%
\bibitem [{\citenamefont {Ach\'{u}carro}\ \emph {et~al.}(2018)\citenamefont
  {Ach\'{u}carro}, \citenamefont {Kallosh}, \citenamefont {Linde},
  \citenamefont {Wang},\ and\ \citenamefont {Welling}}]{Achucarro:2017ing}%
  \BibitemOpen
  \bibfield  {author} {\bibinfo {author} {\bibfnamefont {A.}~\bibnamefont
  {Ach\'{u}carro}}, \bibinfo {author} {\bibfnamefont {R.}~\bibnamefont
  {Kallosh}}, \bibinfo {author} {\bibfnamefont {A.}~\bibnamefont {Linde}},
  \bibinfo {author} {\bibfnamefont {D.-G.}\ \bibnamefont {Wang}}, \ and\
  \bibinfo {author} {\bibfnamefont {Y.}~\bibnamefont {Welling}},\ }\href
  {\doibase 10.1088/1475-7516/2018/04/028} {\bibfield  {journal} {\bibinfo
  {journal} {JCAP}\ }\textbf {\bibinfo {volume} {1804}},\ \bibinfo {pages}
  {028} (\bibinfo {year} {2018})},\ \Eprint {http://arxiv.org/abs/1711.09478}
  {arXiv:1711.09478 [hep-th]} \BibitemShut {NoStop}%
\bibitem [{\citenamefont {Ach\'ucarro}\ \emph
  {et~al.}(2019{\natexlab{b}})\citenamefont {Ach\'ucarro}, \citenamefont
  {C\'espedes}, \citenamefont {Davis},\ and\ \citenamefont
  {Palma}}]{Achucarro:2018ngj}%
  \BibitemOpen
  \bibfield  {author} {\bibinfo {author} {\bibfnamefont {A.}~\bibnamefont
  {Ach\'ucarro}}, \bibinfo {author} {\bibfnamefont {S.}~\bibnamefont
  {C\'espedes}}, \bibinfo {author} {\bibfnamefont {A.-C.}\ \bibnamefont
  {Davis}}, \ and\ \bibinfo {author} {\bibfnamefont {G.~A.}\ \bibnamefont
  {Palma}},\ }\href {\doibase 10.1103/PhysRevLett.122.191301} {\bibfield
  {journal} {\bibinfo  {journal} {Phys. Rev. Lett.}\ }\textbf {\bibinfo
  {volume} {122}},\ \bibinfo {pages} {191301} (\bibinfo {year}
  {2019}{\natexlab{b}})},\ \Eprint {http://arxiv.org/abs/1809.05341}
  {arXiv:1809.05341 [hep-th]} \BibitemShut {NoStop}%
\bibitem [{\citenamefont {Gong}\ and\ \citenamefont
  {Tanaka}(2011)}]{Gong:2011uw}%
  \BibitemOpen
  \bibfield  {author} {\bibinfo {author} {\bibfnamefont {J.-O.}\ \bibnamefont
  {Gong}}\ and\ \bibinfo {author} {\bibfnamefont {T.}~\bibnamefont {Tanaka}},\
  }\href {\doibase 10.1088/1475-7516/2012/02/E01,
  10.1088/1475-7516/2011/03/015} {\bibfield  {journal} {\bibinfo  {journal}
  {JCAP}\ }\textbf {\bibinfo {volume} {1103}},\ \bibinfo {pages} {015}
  (\bibinfo {year} {2011})},\ \bibinfo {note} {[Erratum: JCAP1202,E01(2012)]},\
  \Eprint {http://arxiv.org/abs/1101.4809} {arXiv:1101.4809 [astro-ph.CO]}
  \BibitemShut {NoStop}%
\bibitem [{\citenamefont {Gong}(2016)}]{Gong:2016qmq}%
  \BibitemOpen
  \bibfield  {author} {\bibinfo {author} {\bibfnamefont {J.-O.}\ \bibnamefont
  {Gong}},\ }\href {\doibase 10.1142/S021827181740003X} {\bibfield  {journal}
  {\bibinfo  {journal} {Int. J. Mod. Phys.}\ }\textbf {\bibinfo {volume}
  {D26}},\ \bibinfo {pages} {1740003} (\bibinfo {year} {2016})},\ \Eprint
  {http://arxiv.org/abs/1606.06971} {arXiv:1606.06971 [gr-qc]} \BibitemShut
  {NoStop}%
\end{thebibliography}%

\end{document}